\documentclass[fleqn,usenatbib]{mnras}

\usepackage{newtxtext,newtxmath}
\usepackage[T1]{fontenc}
\usepackage{float}
\usepackage{siunitx}

\DeclareRobustCommand{\VAN}[3]{#2}
\let\VANthebibliography\thebibliography
\def\thebibliography{\DeclareRobustCommand{\VAN}[3]{##3}\VANthebibliography}

\usepackage{graphicx}	
\usepackage{amsmath}	
\usepackage[dvipsnames]{xcolor}
\usepackage{tabularx}
\usepackage{xspace}
\usepackage{verbatim}

\setcitestyle{notesep={}} 

\newcommand{\dcc}{P2200273-v5} 
\newcommand{\tds}{ET-0198D-22} 
\newcommand{\tdsurl}{https://apps.et-gw.eu/tds/?content=3&r=18033} 

\newcommand{\fgw}{f_{\mathrm{GW}}} 
\newcommand{\fgl}{\Delta f_\mathrm{gl}} 
\newcommand{\fdotgl}{\Delta \dot{f}_\mathrm{gl}}
\newcommand{\frot}{f_\mathrm{rot}} 
\newcommand{\fdotrot}{\dot{f}_\mathrm{rot}} 
\newcommand{\flag}{\Delta f_\mathrm{lag}} 

\newcommand{\nglitchATNF}{623\xspace}
\newcommand{\nglitchJodrell}{666\xspace}
\newcommand{\nglitchJodrellmatched}{664\xspace}
\newcommand{\nglitchJodrellunique}{115\xspace}
\newcommand{\nglitchtotal}{740\xspace}
\newcommand{\npulsarstotal}{225\xspace}
\newcommand{\nglitchusable}{726\xspace}
\newcommand{\npulsarsusable}{217\xspace}
\newcommand{\nglitchOfour}{36\xspace}
\newcommand{\nglitchOfive}{74\xspace}
\newcommand{\nglitchOfourYJ}{6\xspace}
\newcommand{\nglitchOfiveYJ}{14\xspace}
\newcommand{\nglitchspd}{627\xspace}
\newcommand{\npulsarsspd}{188\xspace}
\newcommand{\nglitchQATNF}{118\xspace}
\newcommand{\nglitchQ}{119\xspace}
\newcommand{\nglitchQloweps}{73\xspace}
\newcommand{\Oms}{\Omega_\textrm{s}}

\newcommand{\Egw}{E_\mathrm{GW}}
\newcommand{\rwin}{\mathrm{r}}
\newcommand{\ewin}{\mathrm{e}}
\newcommand{\Ntemp}{N_\mathrm{temp}} 

\newcommand{\hs}{\kern-0.9em}
\newcommand{\hsp}{\kern-2em}

\newcommand{\orc}{\includegraphics[height=\fontcharht\font`A]{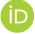}}
\newcommand{\orcid}[1]{\href{https://orcid.org/#1}{\orc}}

\title[Prospects for long transient GWs from glitching pulsars]{Prospects for detecting transient quasi-monochromatic gravitational waves from glitching pulsars with current and future detectors}

\author[J.~Moragues, L.M.~Modafferi, R.~Tenorio, D.~Keitel]{Joan Moragues\,\orcid{0000-0003-2361-2811}\,\thanks{E-mail: joan.moragues@ligo.org},
Luana M. Modafferi\,\orcid{0000-0002-3422-6986}\,\thanks{E-mail: luana.modafferi@ligo.org},
Rodrigo Tenorio\,\orcid{0000-0002-3582-2587}\,\thanks{E-mail: rodrigo.tenorio@ligo.org},
David Keitel\,\orcid{0000-0002-2824-626X}\,\thanks{E-mail: david.keitel@ligo.org}
\\
Departament de F\'isica, Universitat de les Illes Balears, IAC3--IEEC, Crta. Valldemossa km 7.5, E-07122 Palma, Spain
\vspace{-\baselineskip}
}

\date{dated 2022-12-16 -- \href{https://dcc.ligo.org/\dcc}{LIGO-\dcc} -- \href{\tdsurl}{\tds} \\ 
This is a pre-copyedited, author-produced PDF of an article accepted for publication in MNRAS following peer review. The version of record (Monthly Notices of the Royal Astronomical Society, 2022; stac3665) is available online at: \url{https://doi.org/10.1093/mnras/stac3665}}

\pubyear{2022}

\begin{document}
\label{firstpage}
\pagerange{\pageref{firstpage}--\pageref{lastpage}}
\maketitle

\begin{abstract}
Pulsars are rotating neutron stars that emit periodic electromagnetic radiation.
While pulsars generally slow down as they lose energy, some also experience glitches: 
spontaneous increases of their rotational frequency. 
According to several models, these glitches can also lead to the emission of long-duration transient gravitational waves (GWs).
We present detection prospects for such signals
by comparing indirect energy upper limits on GW strain for known glitches
with the sensitivity of current and future ground-based GW detectors.
We first consider the optimistic case of generic constraints based on the glitch size
and find that realistic matched-filter searches
in the fourth LIGO--Virgo--KAGRA observing run (O4)
could make a detection,
or set constraints below these indirect upper limits,
for equivalents of \nglitchOfour out of \nglitchusable previously observed glitches, 
and \nglitchOfive in the O5 run.
With the third-generation Einstein Telescope or Cosmic Explorer, 35--40\% of glitches would be accessible.
When specialising to a scenario where transient mountains produce the post-glitch GW emission,
following the Yim \& Jones model,
the indirect upper limits are stricter.
Out of the smaller set of \nglitchQ glitches with measured healing parameter,
as needed for predictions under that model,
only \nglitchOfourYJ glitches would have been within reach for O4 and \nglitchOfiveYJ for O5,
with a similar percentage as before with third generation detectors.
We also discuss how this model matches the observed glitch population.
\end{abstract}



\section{Introduction}
\label{sec:1}

Pulsars are highly magnetized neutron stars (NSs)
that emit regular pulses of electromagnetic radiation as they spin \citep{LorimerKramer2012}.
In general, their rotation slows down over time due to the energy loss from radiation, 
but sometimes they experience a sudden increase in rotational frequency: 
this phenomenon is known as a \emph{glitch} \citep{Link:1992mdl,Link:2000mu,Lyne:2000sta,main_1}.
During a glitch, a pulsar can suffer not only an increase in its rotational frequency,
but also a change in its spindown parameters (frequency derivatives). 
Glitches can be an ideal probe to understand the physics of NS interiors \citep{Haskell:2017ngx}.

NSs in general are promising sources of gravitational wave (GW) emission \citep{Glampedakis:2017nqy}.
One kind of GWs emitted from NSs are continuous waves (CWs).
These are much fainter than the signals generated by compact binary coalescences \citep{GWTC3}, 
which is the main reason why they remain undetected up to date~\citep{P2200093}.
Improvements of the current LVK detectors --
LIGO Livingston, LIGO Hanford, Virgo and KAGRA \citep{aLIGO, Virgo_detect, KAGRA} --
as well as the new detectors planned for the next decade~\citep{LIGOScientific:2016wof, Adhikari:2019zpy} --
like the Einstein Telescope~\citep[ET,][]{Punturo:2010zz} and Cosmic Explorer~\citep[CE,][]{CE_def} --
give more optimistic predictions for the future \citep{Pitkin_2011, ET_def, Evans:2021gyd}.

In this work we focus on long-duration transient CW-like signals (tCWs) from glitching pulsars.
We do not cover short burst-like signals from $f$-modes,
for which see \citet{Andersson:1997rn, Ho:2020nhi, Yim:2022qcn, Lopez:2022yph}.
Rather, we are interested in tCW signals that are quasi-monochromatic, like CWs,
but with durations of hours to months \citep{Prix_G_M}.
Such long-duration transient signals triggered by pulsar glitches
are predicted by several models \citep{models_1, models_2, Melatos_2015, Singh:2016ilt, Yim_2020}.
The aim of this work is to estimate the detection prospects for such signals 
taking into account sensitivity upgrades to the current detector network and future detectors.
This is complementary to a similar study done for CWs from known pulsars by \citet{Pitkin_2011}.
To do so, we build on past analyses of the population of known glitching pulsars \citep{main_1,main_2} 
using the most recent datasets
of the ATNF and Jodrell Bank glitch catalogues \citep{ATNF_glitches, Jodrell_glitches,Espinoza_2011,Jodrell_extra}.
We use this previously observed glitch population as a proxy for expected future glitches.

Our main detectability results are based on estimating the energy budget available from a glitch,
for example within the two-fluid model for pulsar glitches, 
explained in Section \ref{sec:tCWs}.
If we assume all this energy goes into the emitted GWs,
we get a generic (and optimistic) upper limit for the amplitude of the signal.
For all known glitches, we compare these indirect upper limits
to the expected sensitivity of current and future detectors.
To realistically determine the expected sensitivity of each detector to this particular type of signals,
we consider a specific search method,
the matched-filter transient $\mathcal{F}$-statistic \citep{Prix_G_M},
with realistic setups
as previously used in real searches \citep{Keitel_2019, our_paper, Luana2021}.
For different search methods the sensitivity will vary, 
but this is the method that has already been used in practice,
and it also serves as a limiting case for the sensitivity achievable by other, more generic methods.

The paper is structured as follows:
In Section \ref{sec:obs_pop}, we present a brief overview of the known glitch population.
In Section \ref{sec:tCWs} we briefly review the physics behind pulsar glitches and their GW emission.
Then, in Section \ref{sec:search_methods} we go through some of the search methods for long-duration transient GWs.
We then estimate detection prospects first based on generic indirect energy upper limits (Section \ref{sec:5})
and then more specifically for the \citet{Yim_2020} transient mountain model (Section \ref{sec:mountains}).
In the latter, we also study how well that model matches the observed glitch population.
We conclude in Section \ref{sec:conc}.

\section{Observed population of glitching pulsars}
\label{sec:obs_pop}

Ideally, a study of future detection prospects would be based on a detailed modelling
of the observed population of glitching pulsars
or population synthesis simulations.
We choose a simpler approach,
using the previously observed set of glitches from the known pulsar population, 
assuming that future glitches will have similar characteristics to previous ones,
and leaving a proper treatment of selection effects to future work. 
With the expected larger number of pulsars to be detected by new telescopes,
like CHIME, SKA and FAST \citep{CHIME, SKA, FAST},
our estimates are likely to be pessimistic in terms of the total number of detectable signals.
Conclusions drawn from this study should therefore be a conservative estimation
of the expected number of detectable tCWs emitted by future glitches.
For most glitches,
we take information from
the ATNF and Jodrell Bank glitch catalogues \citep{ATNF_glitches, Jodrell_glitches,Espinoza_2011,Jodrell_extra}.
Both contain lists of glitches observed by various telescopes,
but do not overlap exactly.
In the versions downloaded for this work, the ATNF glitch catalogue includes \nglitchATNF glitches,
and the Jodrell Bank catalogue includes \nglitchJodrell glitches.
Both of them list, for most events, the name of the pulsar, the moment at which it took place 
and the relative changes in frequency and first-order spindown, including error bars.
The ATNF glitch catalogue also contains the ``healing parameter'' $Q$ for \nglitchQATNF glitches,
which indicates the fraction of frequency change that is recovered 
after the glitch \citep{Lyne:2010ad, Haskell:2013goa} 
and the time $\tau$ that it takes for the pulsar to return to a stable state again.
$Q$ ranges from $0$ to $1$,
while typical values for $\tau$ range from a few days to weeks for most pulsars
but can also exceed a year for some of them.
These two last variables are relevant for the \citet{Yim_2020} transient mountain model.

We also work with the main ATNF pulsar catalogue \citep{Manchester:2004bp},
which lists the information for more than 3000 pulsars, such as
name, position, frequency of rotation and different spindown orders.
We access it with the help of the \texttt{psrqpy} package \citep{psrqpy}.
In Appendix \ref{sec:A0} we go through the procedure of matching the three catalogues,
eliminating duplicate glitches,
and identifying cases with missing parameters. 
In the end, we are left with \nglitchtotal unique glitches
from \npulsarstotal pulsars,
of which \nglitchusable from \npulsarsusable have all parameters available
that we need for our main analysis.
These also include two additional glitches collected from the recent literature \citep{UTMOST_2,Zubieta:2022umm}.

Typically, relative changes in rotational frequency $\frot$ and first order spindown $\fdotrot$ are, respectively,
\mbox{$\fgl/\frot \sim 10^{-11} - 10^{-5}$}
and \mbox{$\fdotgl/\fdotrot \sim 10^{-5} - 10^{-2}$}
\citep{main_1}.
It has been noted before that the distribution of absolute glitch sizes
seems to follow a bimodal Gaussian distribution \citep{main_1, main_2, Arumugam:2022ugq}: 
there is a wider component of minor glitches and a narrower one of major glitches (see also Appendix \ref{sec:A2}). 
It is customary to identify these two groups of glitches as ``Crab-like'' and ``Vela-like'', respectively.
\citet{main_2} obtained a value of \mbox{$\fgl = 10^{-5.5}$\,Hz}
as a threshold between the two types of glitches.
The actual sizes of known Crab glitches, however, range from 
approximately $10^{-8}$ Hz to $1.53 \times 10^{-5}$ Hz \citep{Crab_largest}, with two glitches above $10^{-5.5}$.
On the other hand there are 25 observed glitches from the Vela pulsar, and only 19 of them have
\mbox{$\fgl > 10^{-5.5}$\,Hz}.
Hence we will not follow the strict distinction into two subpopulations:
we will use Crab and Vela as prototypes for pulsars with smaller and larger glitches, respectively, 
but without implying that all glitches from the same pulsar have the same properties.

\section{tCWs from pulsar glitches}
\label{sec:tCWs}

\subsection{The basic idea}
\label{sec:tCWs_basics}
Different models have been proposed to describe the underlying physics of NS glitches \citep{Haskell_Melatos}.
A leading idea is that they originate from the layered structure of NS interiors \citep{Lyne:2000sta}.
For this model it is crucial for there to be a superfluid neutron component in the interior.
The measured rotation frequency $\frot$ from electromagnetic observations depends on the evolution of 
the outer crust and the normal (non-superfluid) component of the NS's interior coupled to it.
For both we can assign the angular frequency $\Omega = 2\pi\frot$.
This frequency decreases over time due to the energy lost as emitted radiation,
as measured in pulsar timing.
The superfluid component, however, can continue to rotate at a different angular velocity $\Omega_\mathrm{s}$.
When the difference \mbox{$\Delta\Omega = \Omega_\mathrm{s} - \Omega$}
reaches some critical value, 
an instability takes place and part of the angular velocity of the superfluid component
is transferred to the outer crust.
As a result, the angular velocity of the outer crust suddenly increases and a glitch occurs.
The standard ``two-fluid model'' \citep{Andersson:2002zd, 2004MNRAS.354..101A} for this instability, 
and alternative models, such as ``starquakes'' \citep{1991ApJ...382..587R, Middleditch_J0537}, 
are discussed in more detail in the context of GW emission in \citet{Prix_G_M}.
We will not go further into discussing glitch models here.
Instead we will only summarize below the specific quantities we need to predict GW emission.

First, we consider the frequency evolution at and after the glitch.
When a glitch takes place, we can consider three contributions:
the secular spindown of the pulsar before the glitch,
the near instantaneous change at the glitch epoch,
and often a partial recovery back towards the original frequency.
We can express the total change $\fgl$ in the pulsar's rotational frequency
at the glitch time
as the sum of a transient part and a permanent part,
and similarly with the change $\fdotgl$ in the pulsar's first-order spindown.
Changes in higher-order spindown parameters are only measured for very few glitches
and not included in the catalogues we use here.
Hence, we use the following definition of the change of the frequency of the pulsar due to the glitch:
\begin{equation}
 \label{eq:fgl_YJ}
 \fgl (t) =
 \left\lbrace\begin{array}{lc}
 0,
 & \text{if}~~ t<t_\text{g}\,, \\ 
 \Delta f_{\text{gl,p}}
 +  \Delta \dot{f}_{\text{gl,p}} \Delta t
 +  \Delta f_{\text{gl,t}} \,e^{-\frac{\Delta t}{\tau}},
 & \text{if}~~ t \geq t_\text{g}\,,
 \end{array}\right.
\end{equation}
where 
$\Delta t = t - t_\text{g}$ is the difference between the current time $t$ and the glitch time $t_\text{g}$,
$\tau$ is the recovery time-scale of the glitch,
and $\Delta f_{\text{gl,p/t}}$ are the permanent and transient parts
of the total change in the frequency of the rotating pulsar. 
For the first-order spindown, following \citet{Yim_2020},
only a permanent change $\Delta \dot{f}_{\text{gl,p}}$ is explicitly included;
see Section \ref{sec:mountains-checks} for further details
on how this parametrization relates with their model.

Another relevant variable is the healing parameter $Q$,
which quantifies the fraction of transient change in frequency:
\begin{equation}
\label{eq:Q}
Q = \frac{\Delta f_{\text{gl,t}}}{\fgl}\,.
\end{equation}
If $Q = 1$, the pulsar's frequency $\frot$ from before the glitch is eventually fully recovered,
while if $Q = 0$, the pulsar maintains the frequency acquired at the glitch.
Inferring $Q$ from pulsar timing requires extended post-glitch observations.
Therefore values for $Q$ are only available for a subset of observed glitches.
In particular, for some analyses in this paper we have used
the \nglitchQATNF glitches with $Q$ listed in the ATNF glitch catalogue \citep{ATNF_glitches}
and the $Q$ for one recent Vela glitch from \citet{Zubieta:2022umm}.

A standard emission scenario for persistent CWs \citep{Haskell:2021ljd} is a mass quadrupole,
or ``mountain'',
which emits at $\fgw = 2\frot$.
It is reasonable to expect that a ``transient mountain''
(a deformation that is created at the glitch and dissipates again after a while)
would emit tCW-like signals.
Such a model was explored in a recent work by \citet{Yim_2020},
where the observed change in spindown after a glitch is entirely associated to an extra torque from a transient mountain,
which is equivalent to assuming that the spindown change is purely transient.
This naturally emits GWs on the timescale of observed pulsar glitch recoveries.
We postpone specific constraints for GWs produced by such transient deformations to Section \ref{sec:mountains},
as well as checks of how well the model's assumptions line up with the glitching pulsar population.

Other models propose different physical mechanisms giving rise to a time-varying quadrupolar moment,
such as oscillation modes or circulations inside the NS:
e.g. Ekman flows~\citep{models_1,models_2, Singh:2016ilt}
or Rossby ($r$-)modes~\citep{Andersson, Friedman, chandra1970, FriedmanSchutz, Santiago-Prieto:2012qwb}.
Our main detectability estimation (Section \ref{sec:5}), however,
is agnostic of the details of post-glitch models,
as it only depends on the overall glitch energy budget.\footnote{
$r$-modes are a special case as their dominant GW emission is at approximately 
\mbox{$\fgw \approx 4\frot/3$}~\citep[with relativistic corrections
and dependence on the neutron-star equation of state, see][]{Yoshida, Idrisy}
and hence the detectability estimates for individual glitches will actually be different,
though our procedure still applies identically.
}

\subsection{Indirect energy upper limits}
\label{sec:3.2}
We are interested in an upper bound for the maximum amplitude of a post-glitch tCW signal. 
Here we partially reproduce the derivation given by~\citet{Prix_G_M}, based on the two-fluid model.
The energy budget considered here is different from,
and larger than,
the excess energy liberated in the glitch itself,
which in turn yields upper limits for shorter-duration GW emission~\citep{Ho:2020nhi}.

According to the two-fluid model, glitches are sourced by a frequency mismatch (lag) between
the normal and superfluid components \mbox{$2\pi \flag = \Omega_{\mathrm{s}} - \Omega = \Delta \Omega > 0$}.
More specifically, as discussed in~\citet{Prix_G_M},
the available energy for glitches
is bounded by the excess energy in the superfluid component due to this frequency
mismatch:
\begin{equation}                                                                                                                                                                         
 \label{eq:E_super}                                                                                                                                                                       
 \Delta E_\textrm{s} 
 = \frac{1}{2}\mathcal{I}_\textrm{s}(\Omega_\textrm{s}^2 - \Omega^2) 
 \approx 4\pi^2 \mathcal{I}_\textrm{s} \frot \flag\,,
\end{equation}
where the approximation assumes that~\mbox{$\Delta \Omega$} is negligible with respect to 
$\Omega$. 
Under the assumption that each glitch fully restores corotation between the normal and superfluid components,
which we take throughout this work,
$\Delta E_\textrm{s}$ is the relevant energy budget for post-glitch tCW emission.
A more detailed discussion of this result is given in Appendix~\ref{sec:energy_budget}.

On the other hand, we know that the total energy carried by a quasi-monochromatic GW signal depends
on a (time-dependent) amplitude parameter $h_0 (t)$, the speed of light $c$, 
the gravitational constant $G$, its frequency $\fgw$, the distance from the source $d$ 
and the total emission time $T$:
\begin{equation}
 \label{eq:E_gw}
 \Egw = \frac{2\pi^2c^3}{5G}\fgw^2\,d^2 \int^{T}h_0^2(t)\,\mathrm{d}t \,.
\end{equation}
This is proportional to the square of the dimensionless root-mean-square amplitude $\hat{h}_0$ of a GW signal:
\begin{equation}
 \label{hhat_def}
 \hat{h}_{0}^2 \,T = \int^T h_0^2(t)\,\mathrm{d}t\,.
\end{equation}

To obtain an upper limit, we can combine equations \eqref{eq:E_super} and \eqref{eq:E_gw}
imposing $\fgw=2\frot$.
This way we describe the maximum value of $\hat{h}_0$ in terms of the distance $d$ from the pulsar, 
the moment of inertia $\mathcal{I}$, the duration $T$, 
the glitch size and the rotational frequency:
\begin{equation}
 \label{hhat}
 \hat{h}_{0} \leq \frac{1}{d}\sqrt{\frac{5G}{2c^3}\frac{\mathcal{I}}{T}\frac{|\fgl|}{\frot}}\,.
\end{equation}
where $\mathcal{I}_{\textrm{s}}\,\flag$ has been replaced with $\mathcal{I}\,\fgl$ due to angular momentum
conservation and the assumption of corotation getting restored by the glitch
(see Appendix~\ref{sec:energy_budget}).

If we assume the signal has constant amplitude and lasts for a time $T = \tau_\mathrm{r}$, 
then simply $\hat{h}_{0} = h_{0,\text{r}}$, and the indirect energy upper limit for such a 
``rectangular'' signal becomes
\begin{equation}
\label{eq:up_lim}
 h_{0,\text{r}} \leq \frac{1}{d}\sqrt{\frac{5G}{2c^3}\frac{\mathcal{I}}{\tau_\mathrm{r}}\frac{|\fgl|}{\frot}}\,.
\end{equation}
Given that equations \eqref{eq:E_gw} and \eqref{hhat_def} depend on 
a simple integral of $h_0^2(t)$, if we were to consider an exponentially
decaying signal with decay time of \mbox{$\tau_\ewin = \tau_\rwin$},
the only difference in equations \eqref{eq:up_lim} and \eqref{eq:eps} would be 
an extra multiplicative factor of $\sqrt{2}$.
When presenting our results, we discuss the relevant scaling factor where needed, 
and in most plots we use double axes to show both cases.
See Appendix~\ref{sec:app_windows} for an explicit calculation.

Assuming that tCW emission is sourced by a transient quadrupolar moment induced 
in the star due to an unspecified process caused by the glitch, 
upper limits on $h_0$ can be interpreted using the standard expression for the 
GW amplitude from a non-axisymmetric deformation~\citep{Jaranowski_1998},
\begin{equation}
 \label{eq:h0_eps_d}
 h_{0}(t) = \frac{4\pi^2G\mathcal{I}}{c^4}\frac{\fgw^2}{d}\epsilon(t)\,.
\end{equation}
Here we use the equatorial ellipticity defined from the principal moments of inertia as 
$\epsilon = \lvert \mathcal{I}_x - \mathcal{I}_y \rvert / \mathcal{I}$, 
where for consistency with the above equations we use a simple $\mathcal{I}$ for the $z$ component.
By equating \eqref{eq:h0_eps_d} to equation \eqref{eq:up_lim} we obtain
\begin{equation}
 \label{eq:eps}
 \epsilon_\text{r} \leq \frac{1}{16\pi^2} \sqrt{\frac{5c^5}{2\,G}
 \frac{1}{\mathcal{I}\,\tau_\text{r}}\frac{|\fgl|}{\frot^5}}\,.
\end{equation}
This provides a benchmark that indicates the maximum possible ellipticity 
a pulsar can present due to a transient mountain formed at a glitch, 
in addition to any long-term ellipticity that it may have. 
It is independent of the pulsar distance and only depends 
on the intrinsic properties of the pulsar and the glitch.
In this picture,
a rectangular signal corresponds to assuming that a transient mountain forms
at the glitch, remains at constant size for time $\tau_\rwin$, 
and disappears instantaneously.

A more specific model for the effects of a transient quadrupolar moment is given by~\citet{Yim_2020},
according to which the source of tCW emission is a transient mountain
built up during or just after the glitch by an unspecified process.
We postpone an in-depth discussion of their model to Section~\ref{sec:mountains-uls}.
Regarding energy balance, their argument is that such a transient mountain
causes a radiative loss which directly impacts the rotational frequency
of the observed pulsar and, thus, can be measured from electromagnetic observations.
More specifically, only a fraction \mbox{$Q$} of the energy transferred 
to the normal component is implied to be radiated.
Consequently, the available budget to be emitted via tCWs under the model
is reduced to \mbox{$E_\mathrm{tCW} \approx Q \Delta E_{\textrm{s}}$}.
This also reduces upper limits from equations~\eqref{hhat},~\eqref{eq:up_lim} 
and~\eqref{eq:eps} by a factor~$\sqrt{Q}$.

\section{Search methods for tCWs}
\label{sec:search_methods}

In this section we present a brief description of different methods to search for tCWs. 
Understanding this is crucial for a meaningful detectability estimation,
as the sensitivity of a search depends not only on the detector noise curves 
but also on the search method and the specific setup.
We concentrate on searches for signals from known pulsars,
though some of the methods could also conceivably be extended to broader searches for unknown sources.

\subsection{Transient $\mathcal{F}$-statistic}
\label{sec:transF}

A commonly used method in CW detection is the $\mathcal{F}$-statistic
\citep{Jaranowski_1998, Cutler}.
It is applicable to many kinds of searches \citep[see][ for an overview]{Tenorio:2021wmz}, 
but here we focus on the variant for tCWs introduced by \citet{Prix_G_M}.
This assumes that the signal is equivalent to a truly persistent CW multiplied with a window function
that limits its duration and modifies its amplitude evolution:
\begin{equation}
 \label{eq:signalmodel}
 h(t;\theta) = h(t;\lambda,\mathcal{A},\mathcal{T}) = \omega(t;\mathcal{T})h(t;\lambda,\mathcal{A})\,.
\end{equation}
Here $\theta$ are all the parameters the signal depends on.
$h(t;\lambda,\mathcal{A})$ is a standard CW described by the amplitude parameters 
\mbox{$\mathcal{A}=\{h_0, \iota,\psi,\phi_0\}$}
(nominal amplitude, inclination angle, polarization angle and initial phase)
and phase evolution parameters
\mbox{$\lambda=\{\fgw^{(k)},\alpha, \delta\}$} \citep{Jaranowski_1998}. 
These include a certain number of spindown terms $\fgw^{(k)}$ for the intrinsic frequency evolution
given by a Taylor expansion,
and a GW detector measures an additional time-dependent Doppler shift
depending on the sky position $(\alpha, \delta)$.

The transient window function $\omega$ depends on a set of transient parameters
\mbox{$\mathcal{T} = \{t_0, \tau \}$},
with $t_0$ the start time and $\tau$ a duration parameter. 
A rectangular window simply turns the signal on at time $t_0$,
and turns it off after a duration $\tau_\mathrm{r}$,
and the amplitude \mbox{$h_0(t)=h_{0,\rwin}$} is constant in between.
A more natural expectation for dissipative processes,
as discussed in Section \ref{sec:tCWs},
is an exponential decay window
\mbox{$\omega(t;\mathcal{T}) = \exp[-(t-t_0)/\tau_\mathrm{e}]$ for $t\geq t_0$},
for which we denote the initial amplitude as
\mbox{$h_0(t_0)=h_{0,\ewin}$}.

Given observed data $x$, we compare two hypotheses: 
$\mathcal{H}_\mathrm{G}$ states that the observed data consists only of Gaussian stationary noise, 
while $\mathcal{H}_\mathrm{tS}$ assumes that
it also contains a tCW signal $h(t;\theta)$.
The likelihood for the data $x$ in the Gaussian noise case is
$P(x|\mathcal{H}_\mathrm{G})$
and the likelihood with a signal is
$P(x|\mathcal{H}_\mathrm{tS})$.
From the ratio
\mbox{$\Lambda(x,\theta) =P(x|\mathcal{H}_\mathrm{tS},\theta)/ P(x|\mathcal{H}_\mathrm{G})$},
one obtains:
\begin{equation}
 \log \Lambda(x;\theta)
 = \mathcal{A}^\mu x_\mu(\lambda,\mathcal{T})
 - \frac{1}{2}\mathcal{A}^\mu\mathcal{M}_{\mu\nu}(\lambda, \mathcal{T})\mathcal{A}^\nu \,,
\end{equation}
where $\mathcal{A}^\mu$ is a re-parametrization \citep{Jaranowski_1998}
of the four amplitude parameters, 
$x_\mu$ are the projections of the data $x$ on the basis functions associated with these amplitudes,
and $\mathcal{M}_{\mu\nu}$ is the antenna pattern matrix. 
Summation over up/downstairs indices is implied.

The $\mathcal{F}$-statistic is the result of introducing the maximum-likelihood estimate
into the log-likelihood ratio:
\begin{equation}
 2\mathcal{F}(x;\lambda, \mathcal{T}) = x_\mu\mathcal{M}^{\mu\nu}(\lambda, \mathcal{T})x_\nu \,.
\end{equation}
So we are left with a detection statistic only dependent on the $\lambda$ and $\mathcal{T}$ parameters.
The CW case corresponds to evaluating this with a rectangular window over the whole observation time.
On the other hand, for tCWs, \citet{Prix_G_M} introduced two detection statistics: 
one is $\max\mathcal{F}$, the maximum $\mathcal{F}$ over transient parameters, 
and the other is a Bayes factor $\mathcal{B}_\mathrm{tS/G}$ obtained by marginalizing over them,
which is slightly more sensitive in Gaussian noise and more robust in real data~\citep{Tenorio:2021wad}.

In either case this method is fully-coherent over the duration of the signal,
meaning that we have to assume a certain phase evolution $\lambda$.
If this is not exactly known, we have to search a template bank of different $\lambda$
covering any uncertainties in these parameters.
Search sensitivity then also depends on the size of the parameter space covered,
as more templates lead to a higher statistical trials factor
and an increased threshold is needed to keep a fixed false-alarm probability~\citep{Tenorio:2021wad}.

The transient-$\mathcal{F}$-statistic method has been used in searches of data from the O2 run \citep{Keitel_2019}
and of O3 data \citep{our_paper}, with template bank sizes up to $\sim 10^7$.
Both analyses used only rectangular search windows, for reasons of efficiency:
Evaluating the tCW detection statistics for exponential windows
would be computationally significantly more expensive \citep{Prix_G_M,Keitel_Ashton_2018},
while using rectangular search windows for exponentially decaying signals
gives only a small loss in detection efficiency \citep{Prix_G_M,Keitel_2019}.

\subsection{Other methods}

While the transient-$\mathcal{F}$-statistic method relies on specific templates for the GW frequency evolution,
other methods make much fewer assumptions and can be used to search for more generic signals.
The Stochastic Transient Analysis Multidetector Pipeline
\citep[STAMP,][]{Thrane:2010ri,stochtrack, STAMP, Macquet:2021ttq}
is an example of such an unmodelled method.
It is based on pattern recognition of time-frequency spectrogram data,
and can be used in several configurations for signals of varying duration,
without imposing a fixed starting time or specific frequency evolution.
This makes it computationally cheaper and more robust towards a wider class of GW signals,
but less sensitive than the matched-filter $\mathcal{F}$-statistic
for signals that actually do follow the tCW model.

Another approach for identifying GW signals while allowing for flexibility in the time-frequency tracks
is based on the Viterbi algorithm \citep{Viterbi:1967}
in combination with a hidden Markov model \citep[HMM,][]{Suvorova:2016rdc}.
It finds the most probable sequence of `hidden states'
(the intrinsic $\fgw(t)$ of a source)
from the observed data.
This can be applied on different time-frequency resolved input statistics,
for example outputs of STAMP \citep{HMM_Viterbi_bonagiri},
raw strain power~\citep{Sun:2018owi},
or the $\mathcal{F}$-statistic evaluated over short segments of data~\citep{HMM_Viterbi_Fstat}.
Even in the latter case,
it is a \textit{semi-coherent} method as opposed to
the fully-coherent transient $\mathcal{F}$-statistic,
because it does not preserve phase information across segments.
SOAP~\citep{2019PhRvD.100b3006B} is another Viterbi-based method
working on Fourier power which has recently
been used to conduct all-sky searches
for quasi-monochromatic gravitational-wave signals~\citep{LIGOScientific:2022pjk}.

Both STAMP and Viterbi have been used before for a variety of searches
\citep[e.g.,][]{interm_STAMP, Abbott_2019_2_4, KAGRA:2021bhs, O3-5}
including long-duration transient searches for a putative NS remnant
of the binary merger GW170817.
But while they should be applicable to tCWs from glitching pulsars too,
this has not been done in practice yet.

Another semi-coherent search method was developed by \citet{Keitel_2015},
again based on the $\mathcal{F}$-statistic.
This was devised as a computationally efficient `add-on' for semi-coherent CW searches,
evaluating the likelihood for tCWs that happen to match closely enough
the fixed duration of segments in that search setup.
It has been applied in the Einstein@Home framework \citep{ET_home},
but no observational results from it have been reported to date.

The fully-coherent $\mathcal{F}$-statistic is computationally too expensive
to extend beyond the small parameter spaces of known pulsars considering
observations of more than a few months.
On the other hand, STAMP, Viterbi-based and other semi-coherent methods
could in principle also be used to search for signals from unknown sources,
or where the glitch has not been noticed in electromagnetic observations
due to insufficient observational coverage.

New approaches with even greater computational efficiency and
the ability to generalise to flexible signal models
can also be expected to become available thanks to novel machine learning techniques.
See \citet{Cuoco} for a review of such techniques in the GW field.
For example, deep learning to search for tCWs from glitching pulsars 
is studied by \citet{Modafferi_CNNs}.

\section{Observational prospects from generic energy upper limits}
\label{sec:5}

Detecting (t)CWs, or at least putting physically interesting constraints on potential sources,
requires improvements in sensitivity over past observing runs.
These are expected from the LVK detectors during the O4 and O5 runs \citep{Virgo_prosp}
and more so with the proposed next-generation detectors like ET and CE.

\begin{figure*}
 \includegraphics[width=\textwidth]{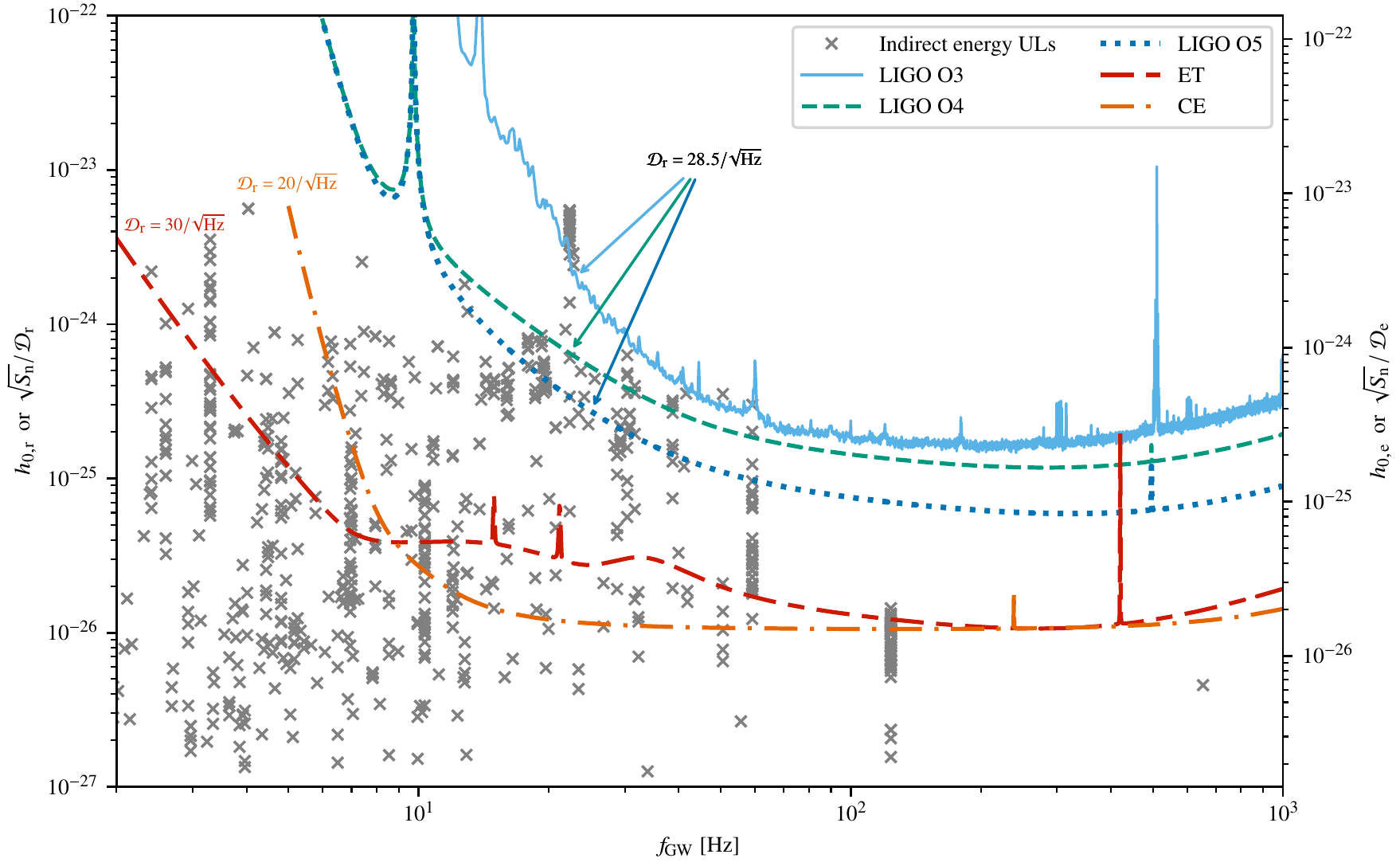}
 \caption{Prospects for detecting or constraining tCWs with current and future detectors,
  extrapolated from previously observed pulsar glitches.
  Plotted against GW frequency,
  the crosses indicate the indirect energy upper limits on GW strain amplitude from a glitch,
  using the total superfluid excess energy and not assuming any particular emission model.
  Curves give the expected sensitivity for realistic matched-filter searches
  with different detector configurations
  in terms of $\sqrt{S_\mathrm{n}}/\mathcal{D}_{\rwin/\ewin}$
  (detector noise curve divided by sensitivity depth of a search).
  For a given glitch,
  tCWs could be detected or physically constrained with a given detector
  when the cross is above the corresponding curve.
  The left-hand vertical axis gives results for ``rectangular window'' signals (constant amplitude)
  directly from equation \eqref{eq:up_lim}.
  The right-hand axis is for exponentially decaying signals,
  where both the upper limits and the sensitivity curves scale with the same factor of $\sqrt{2}$.
  The depth for each detector (network) configuration takes into account
  the number of detectors and their geometry.
  This figure illustrates the case of
  \mbox{$\tau_{\rwin/\ewin} = 10$\,days},
  but can be easily rescaled to different duration parameters,
  as both the indirect energy upper limits and sensitivity curves
  scale with $1/\sqrt{\tau_{\rwin/\ewin}}$.
  Note that the horizontal axis is \mbox{$\fgw = 2\frot$},
  so that e.g. the stack of crosses near 22\,Hz corresponds to Vela
  and the one near 60\,Hz corresponds to the Crab.
 \label{fig:glitch_vs_psd}
}
\end{figure*}

Current detectors are modified L-shaped Michelson interferometers with arm lengths of
4\,km (LIGO) and 3\,km (Virgo and KAGRA).
ET will consist of 3 detectors in shared tunnels with lengths of 10\,km 
and arm opening angles of 60 degrees \citep{ET_sens_1}.
CE is planned as a single 40\,km L-shaped detector
with an optional second detector of 20\,km or 40\,km \citep{Evans:2021gyd}.
Both concepts also include other improvements in
mirrors, lasers, cryogenics, quantum technologies and noise-mitigation measures.
They are planned to be functional in the 2030s.

We will first discuss, in this section, the most generic
and hence also least constraining upper limits,
without specific model assumptions for the post-glitch tCW emission, 
based solely on the available superfluid excess energy
as derived in Section \ref{sec:3.2}.
Then in Section \ref{sec:mountains} we will specialise to
the transient mountain model by \citet{Yim_2020},
for which upper limits are lower,
but which allows us to derive additional constraints on
the distance reach of searches and the ellipticities that can be probed.

The generic results are summarized in Fig.~\ref{fig:glitch_vs_psd}.
This compares indirect energy upper limits for the previously observed glitch population
to the estimated sensitivity of realistic matched-filter searches with different detector configurations, 
both as functions of frequency,
and for an example duration parameter of \mbox{$\tau = 10$\,days}.
The energy upper limits on $h_0$ are computed applying equation \eqref{eq:up_lim}
to data from the ATNF and Jodrell Bank catalogues, 
assuming GW emission at $\fgw = 2 \frot$ extrapolated to a reference time
corresponding to the O4 run starting time.
(Here we analyse \nglitchusable glitches with epoch, $\fgl$, distance and $\frot^{(k)}$
available from the catalogues, see Appendix \ref{sec:A0}.)
The frequency-dependent sensitivity of searches with different detector configurations
is derived from their projected noise curves and a realistic matched-filter search setup
in such a way that,
for a given glitch, tCWs could be detected 
(or at least physically constrained) 
if its marker lies above the corresponding sensitivity curve.

More precisely, these \emph{sensitivity curves} correspond to detectable amplitudes within each frequency band,
averaged over some signal population,
given a specific search setup, noise curve and amount of data;
e.g. $h_0^{95\%}$ at a confidence level of 95\%.\footnote{
This quantity is usually referred to as ``population-based upper limits''
when CW search results are presented;
see e.g. \citet{Tenorio:2021wmz}.
Also see \citet{Wette_2012} for details on population averaging procedures.
}
To construct these conveniently for different detector configurations,
we re-scale the detector noise amplitude spectral densities (ASDs) $\sqrt{S_\textrm{n}}$
by the \textit{sensitivity depth} \citep{Behnke:2014tma,Dreissigacker2018} defined as
(omitting for simplicity the confidence level on its symbol)
\begin{equation}
 \label{eq:depth}
 \mathcal{D} = \frac{\sqrt{S_\textrm{n}}}{h_0^{95\%}} \,.
\end{equation}

Briefly, this states
how deep under the noise we expect to be able to detect a population of signals,
using a specific search configuration.
Hence, larger $\mathcal{D}$ implies a more sensitive search.
In search setups with fixed false-alarm probability per frequency band,
there is a single depth across all frequencies.
As we see below, that is also true to good approximation for the search setups we consider.
We can then apply such a single factor
for pulsars at different frequencies and for different observing runs of a given detector network.
We now discuss how this factor is estimated and scales with various parameters
before proceeding to the interpretation of results.

First, just like the indirect upper limits
(see equation \eqref{eq:up_lim}),
sensitivity depth depends on the amplitude evolution window and the duration of tCW signals.
In particular, for a tCW search with the transient $\mathcal{F}$-statistic
(as discussed in Section~\ref{sec:transF}),
the sensitivity depth for rectangular signals of duration $\tau_\rwin$ is similar to a fully-coherent
CW search of that duration \citep{Keitel_2019},
meaning it scales with $\sqrt{\tau_\rwin}$~\citep{Jaranowski_1998, P2200093}.
As derived in Appendix \ref{sec:app_windows}, the difference between rectangular and exponential 
amplitude evolution windows with \mbox{$\tau_\rwin=\tau_\ewin$}
is simply an overall factor $\sqrt{2}$ in both
upper limits
and sensitivity curves ($\sqrt{S_\mathrm{n}}/\mathcal{D_{\rwin/\ewin}}$).
This scaling is taken into account in the secondary vertical axis in Fig. \ref{fig:glitch_vs_psd},
and scalings with $\tau_{\rwin/\ewin}$ remain the same for both cases.
In summary, this implies that the relative detectability of signals with different $\tau_{\rwin/\ewin}$ remains the same.\footnote{
This is still true when there are gaps in the data,
as long as they are small compared to the $\tau_{\rwin/\ewin}$ under consideration and approximately uniformly distributed.
There will only be an overall reduction in sensitivity given by the overall duty factor.
A counter-example where the scaling of sensitivity slows down notably was the one-month maintenance break in O3.
}
Fig. \ref{fig:glitch_vs_psd} illustrates the case of
\mbox{$\tau_{\rwin/\ewin} = 10$\,days},
but can be easily rescaled to different durations.
This assumes that the transient $\mathcal{F}$-statistic search
uses a window function matching the actual signal shape.
If a real search uses only the computationally cheaper rectangular window
but the real signal decays exponentially, 
there will be some mild sensitivity loss \citep{Prix_G_M, Keitel_2019}.

To cover the remaining dependencies
(population parameters,
statistical properties of the search method,
search setup,
and the average duty factor, location and geometry of the detectors),
we use the numerical sensitivity depth estimation procedure from \citet{Dreissigacker2018},
as implemented in \texttt{OctApps} \citep{Wette_2018}.
One of the inputs needed for this calculation is a threshold on the detection statistic.
Our threshold is obtained by first simulating noise-only transient $\mathcal{F}$-statistic
samples matching realistic setups \citep{Keitel_2019, our_paper, Luana2021}
for narrowband searches (\mbox{$\fgw\approx2\frot$})
on glitches from Crab and Vela.
We then estimate the distribution for the loudest outlier
using the method from~\citep{Tenorio:2021wad}
as implemented in the \texttt{distromax} package~\citep{distromax}.
Details on this procedure are included in Appendix \ref{sec:depth}.

As discussed above, the depth is independent of the detector noise curve,
but it still depends on the number of detectors involved and their geometry.
We now discuss the specific detector (network) configurations
included in Fig.~\ref{fig:glitch_vs_psd},
and the sensitivity depths we obtain.

First, we represent the LVK network by just the two LIGO detectors,
which we consider as approximately equally sensitive,
as also required by the \texttt{OctApps} implementation.
For the last completed observing run O3 we use
the harmonic mean of the ASDs from LIGO Hanford and LIGO Livingston
taken from \citet{O3_PSD_H1, O3_PSD_L1}.
For the upcoming runs O4 and O5 we use the nominal ASDs from \citet{O4_O5}.
We do not add the less sensitive Virgo and KAGRA detectors~\citep{Virgo_prosp, KAGRA},
so the resulting estimate will be somewhat conservative,
though their inclusion in practical searches also depends on computational cost considerations.
We find sensitivity depths for Crab and Vela searches of
$\mathcal{D}_\rwin \approx 27.5$
and \mbox{$29.5/\sqrt{\mathrm{Hz}}$}, respectively,
considering a rectangular signal with
\mbox{$\tau_\rwin = 10$\,days}.
Differences are due to the two pulsars' distinct sky locations
and the different number of templates needed in each search setup,
leading to different detection thresholds
(see Appendix \ref{sec:depth}).
As we will see later in this section,
that difference is small enough that we can simplify to
using the average
\mbox{$\mathcal{D}_\rwin = 28.5/\sqrt{\mathrm{Hz}}$}
for all pulsars,
independent of sky position and frequency.

Similarly, for ET and CE we use average depths from the Crab and Vela example setups.
But the network configurations are different, leading to different depth values.
For a single CE detector with the same L-shape geometry as LIGO,
assumed to be located in Hanford
and with ASD taken from \citet{T1500293},
the depth is similar to a single LIGO detector,
\mbox{$\mathcal{D}_\rwin \approx 20/\sqrt{\mathrm{Hz}}$}.
For ET, we assume it to be located at the Virgo site.
We start from the ASD~\citep{ET-0000A-18} of a single detector
and then estimate the depth including the full triangular observatory geometry;
see appendix \ref{sec:depth}.
The resulting depth is \mbox{$\mathcal{D}_\rwin \approx 30 / \sqrt{\mathrm{Hz}}$}.

\begin{figure}
 \includegraphics[width=\columnwidth]{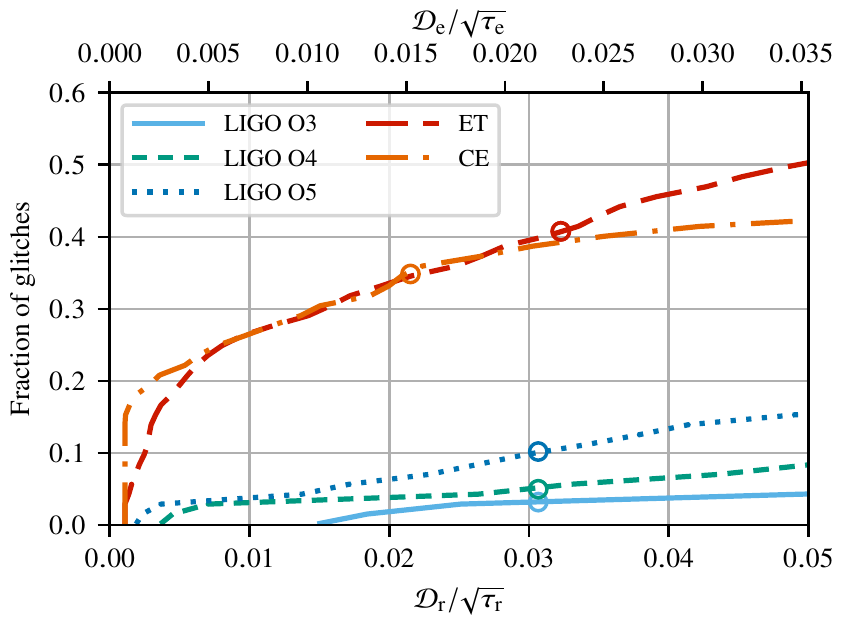}
 \caption{
  Fraction of previously observed glitches
  whose tCW emission could be detected or physically constrained
  with a given detector
  as a function of search sensitivity depth rescaled by signal duration,
  using the total superfluid excess energy
  and not assuming any particular emission model.
  The circles on each curve represent the depth factors
  as used in Fig. \ref{fig:glitch_vs_psd},
  rescaled by $\tau=10$\,days,
  for each detector configuration.
  The lower horizontal axis is for rectangular window signals,
  while the upper one is for exponential windows.
  Detectability is the same for both cases,
  because the scaling of the indirect energy upper limits
  cancels with that of the depth.
 \label{fig:stair_plot}
}
\end{figure}

The estimated depths just discussed were used for the sensitivity curves
in Fig.~\ref{fig:glitch_vs_psd}.
On the other hand, in Fig.~\ref{fig:stair_plot}
we count how many previously observed glitches
would have indirect energy upper limits above the search sensitivity curve,
with each detector configuration,
when varying the depths. 
More precisely,
we plot against $\mathcal{D}/\sqrt{\tau}$
to remove the dependency on the signal duration. 
This quantity depends solely on the search method and setup.
The figure demonstrates that taking into account a realistic depth for a given search
is essential
for a fair assessment of detection prospects.
On the other hand, small changes in depth do not change the result significantly:
the slopes near our estimated values are relatively shallow.
This justifies taking the average between Crab and Vela results
as approximately representative for all pulsars.
The same figure also demonstrates that different detector configurations may reach
the physically constraining regime for a very different number of pulsar glitches.
This will also be the case for different search methods if their depths differ by more than a few, 
and at least a qualitatively accurate sensitivity estimation for each approach 
is crucial in discussing detection prospects.
As mentioned in Section~\ref{sec:search_methods},
we generally expect other, more generic search methods
to be less sensitive than the transient $\mathcal{F}$-statistic,
as long as the signals follow the signal model of \citet{Prix_G_M},
but they would still be highly valuable to cover more generic signal types.

\begin{table}
 \centering
 \caption{
  All pulsars for which a transient $\mathcal{F}$-statistic search
  in the O4 or O5 runs of the LVK detectors 
  could reach below the indirect energy upper limits
  from equation \eqref{eq:up_lim}
  for at least one of their previous glitches,
  using the total superfluid excess energy
  and not assuming any particular emission model. 
  As discussed in the main text,
  these results are independent of signal duration.
  The first three columns list
  the pulsar name,
  its approximate GW frequency assuming dominant mass quadrupole emission,
  and the upper limit for the largest glitch seen from this pulsar,
  assuming a rectangular amplitude window.
  The last two columns state
  how many of the previously observed glitches from each pulsar
  have such an upper limit above the sensitivity curve
  from O4 and O5 respectively,
  and the corresponding percentage.
  \label{tab:1}
 }
\begin{tabular}{ccccrcr}
\hline
 PSR   &  $\fgw$\,[Hz]  \hs &  max $h_{0,\rwin}$ \hs  & \multicolumn{2}{c}{O4\hs} &  \multicolumn{2}{c}{O5}   \\
\hline
J0205+6449 &    30.4     \hs&  \num{6.3e-25} \hs   &  3/9 \hs & \hs (33 \%) \hs &  4/9 \hs & \hs (44 \%)  \\
J0358+5413 &    12.8     \hs&  \num{1.8e-24} \hs  &  0/6 \hs & \hs (0 \%) \hs &  1/6 \hs & \hs (17 \%)  \\
J0534+2200 &    59.2     \hs&  \num{1.2e-25} \hs   & 2/30 \hs & \hs (7 \%) \hs & 3/30 \hs & \hs (10 \%)  \\
J0835-4510 &    22.4     \hs&  \num{4.7e-24} \hs  & 21/25 \hs & \hs (84 \%) \hs & 23/25 \hs & \hs (92 \%) \\
J0940-5428 &    22.8     \hs&  \num{2.9e-24} \hs  &  2/2 \hs & \hs (100 \%) \hs & 2/2 \hs & \hs (100 \%)  \\
J1023-5746 &    17.9     \hs&  \num{7.9e-25} \hs   &  0/7 \hs & \hs (0 \%) \hs &  6/7 \hs & \hs (86 \%)  \\
J1028-5819 &    21.9     \hs&  \num{9.2e-25} \hs   &  1/1 \hs & \hs (100 \%) \hs & 1/1 \hs & \hs (100 \%)  \\
J1105-6107 &    31.6     \hs&  \num{1.9e-25} \hs   &  2/6 \hs & \hs (33 \%) \hs &  4/6 \hs & \hs (67 \%)  \\
J1112-6103 &    30.8     \hs&  \num{2.6e-25} \hs   &  0/4 \hs & \hs (0 \%) \hs & 4/4 \hs & \hs (100 \%)  \\
J1413-6205 &    18.2     \hs&  \num{5.3e-25} \hs   &  0/1 \hs & \hs (0 \%) \hs & 1/1 \hs & \hs (100 \%)  \\
J1420-6048 &    29.3     \hs&  \num{2.2e-25} \hs   &  0/7 \hs & \hs (0 \%) \hs &  2/7 \hs & \hs (29 \%)  \\
J1524-5625 &    25.6     \hs&  \num{4.4e-25} \hs   &  0/1 \hs & \hs (0 \%) \hs & 1/1 \hs & \hs (100 \%)  \\
J1531-5610 &    23.7     \hs&  \num{5.0e-25} \hs    &  0/1 \hs & \hs (0 \%) \hs  & 1/1 \hs & \hs(100 \%)  \\
J1617-5055 &    28.8     \hs&  \num{2.6e-25} \hs   &  0/7 \hs & \hs (0 \%) \hs &  1/7 \hs & \hs (14 \%)  \\
J1709-4429 &    19.5     \hs&  \num{4.8e-25} \hs   &  0/6 \hs & \hs (0 \%) \hs &  5/6 \hs & \hs (83 \%)  \\
J1809-1917 &    24.2     \hs&  \num{3.4e-25} \hs   &  0/1 \hs & \hs (0 \%) \hs & 1/1 \hs & \hs (100 \%)  \\
J1813-1246 &    41.6     \hs&  \num{3.6e-25} \hs   &  1/1 \hs & \hs (100 \%) \hs & 1/1 \hs & \hs (100 \%)  \\
J1826-1334 &    19.7     \hs&  \num{4.6e-25} \hs   &  0/8 \hs & \hs (0 \%) \hs &  1/8 \hs & \hs (12 \%)  \\
J1907+0602 &    18.7     \hs&  \num{7.9e-25} \hs   &  0/2 \hs & \hs (0 \%) \hs &  1/2 \hs & \hs (50 \%)  \\
J1952+3252 &    50.6     \hs&  \num{3.5e-25} \hs   &  1/6 \hs & \hs (17 \%) \hs &  1/6 \hs & \hs (17 \%)  \\
J2021+3651 &    19.3     \hs&  \num{7.8e-25} \hs   &  0/5 \hs & \hs (0 \%) \hs &  4/5 \hs & \hs (80 \%)  \\
J2229+6114 &    38.7     \hs&  \num{3.1e-25} \hs   &  3/9 \hs & \hs (33 \%) \hs &  6/9 \hs & \hs (67 \%)  \\
\hline
\textbf{TOTAL} & - \hs& - \hs& \color{white}i\color{black}36/726\hs &\hs(5 \%) \hs& \color{white}i\color{black}74/726\hs &\hs(10 \%)
\end{tabular}
\end{table}

In summary, we find that LIGO (or the full LVK network) at design sensitivity
should already be able to put physical constraints on tCWs from a small fraction of glitches,
reaching below the generic indirect energy upper limits.
But prospects become significantly better with the next generation of detectors,
with both CE and ET able to constrain or detect tCWs
from a large fraction of the glitch population.
To guide searches in the two upcoming LVK runs O4 and O5,
we also provide more detailed results in Table \ref{tab:1}.
This lists all pulsars for which at least one of its previous glitches
would have been accessible at O4 and O5 sensitivities,
assuming a depth of
\mbox{$\mathcal{D}_\rwin=28.5/\sqrt{\mathrm{Hz}}$}.
There are 9 such pulsars for O4 and 22 for O5. 
Vela (J0835$-$4510),
in particular,
is a pulsar of interest for tCW searches
not only because it glitches frequently, 
but also because the majority of its glitches could be detectable in the near future.
From the \nglitchOfour previous glitches from any pulsars accessible at O4 sensitivity,
21 are from Vela
(23 out of \nglitchOfive for O5).
The Crab (J0534+2200),
on the other hand,
has 30 known glitches,
but only 2 of them have been large enough for O4 sensitivity (3 for O5).
Unfortunately, the ``big glitcher'' J0537$-$6910 \citep{Middleditch_J0537, Ho_J0537},
according to these estimates,
is not within reach for O4 nor O5 searches,
mostly due to its large distance.
Even so,
4 of its known past glitches would have been accessible with the expected ET sensitivity.

\section{Transient mountain scenario}
\label{sec:mountains}

The indirect upper limits discussed in Section~\ref{sec:5}
only assume the overall energy budget of a glitch triggered by two-fluid instabilities and,
as discussed by \citet{Prix_G_M}, they also hold approximately for different models, e.g. starquakes.
They are therefore independent of specific models for the post-glitch phase:
as long as the available energy is fixed, and completely converted into GWs,
the details of this conversion process do not matter for these upper limits.
Now we will make a more specific assumption:
the tCW signal is emitted by a transient deformation (``mountain'')
formed at the time of the glitch.

We will first take the phenomenological model proposed by \citet{Yim_2020}
for this scenario at face value and
re-estimate observational prospects
(Section \ref{sec:mountains-uls})
as well as studying the additional constraints on source parameters
that could be obtained under the model
(Section \ref{sec:mountains-constraints}).
In Section \ref{sec:mountains-checks},
we further investigate how consistent the model is with the observed glitch population.
We have briefly introduced the model in Section \ref{sec:tCWs}.
To recap, they consider a transient mountain formed on the pulsar surface
at the moment of the glitch, without specifying the formation process.
Then, this mountain would dissipate over time during the relaxation phase after the glitch
and it would create a braking torque on the NS spinning it back down to near pre-glitch spin
frequencies via the emission of tCWs.

\subsection{Observational prospects under Yim \& Jones model}
\label{sec:mountains-uls}

\begin{figure}
 \includegraphics[width=\columnwidth]{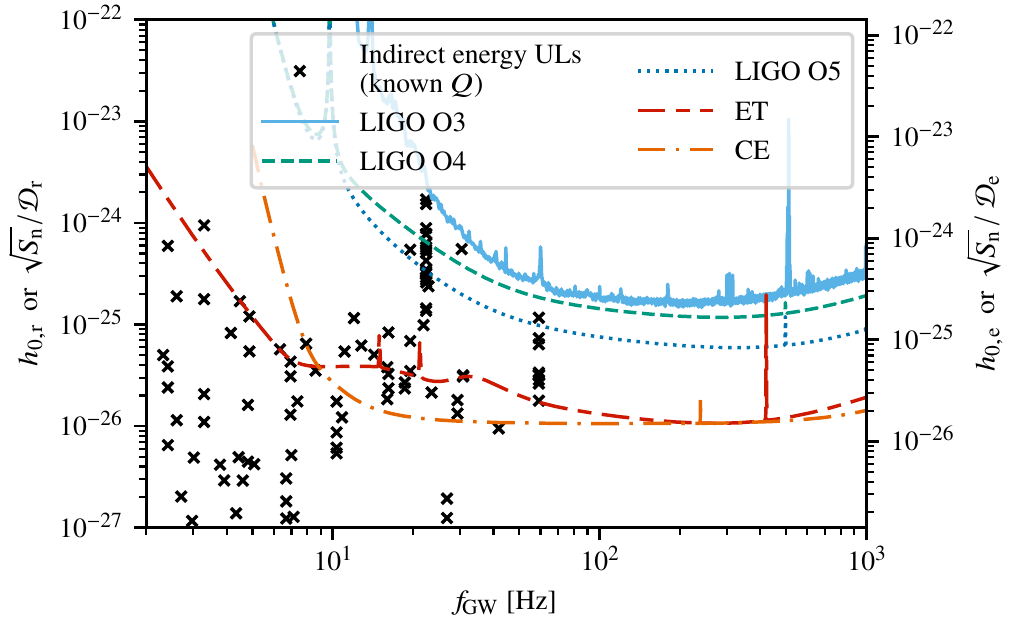}
 \caption{
  Prospects for detecting or constraining tCWs with current and future detectors,
  extrapolated from previously observed pulsar glitches,
  as in Fig.~\ref{fig:glitch_vs_psd};
  but now assuming the transient mountain model of \citet{Yim_2020}
  and limited to \nglitchQ glitches with known healing parameter $Q$.
  The crosses indicate the indirect energy upper limits on GW strain amplitude from a glitch, 
  using equation \eqref{eq:up_lim_Q},
  which are lower than the more generic upper limits in Fig. \ref{fig:glitch_vs_psd}
  by a factor $\sqrt{Q}$.
  As in Fig. \ref{fig:glitch_vs_psd},
  this is calculated for a nominal duration parameter
  \mbox{$\tau_{\rwin,\ewin}=10$\,days}.
  \citet{Yim_2020} predict that a tCW signal from a glitch
  with observed recovery parameter $\tau$ should have
  \mbox{$\tau_\ewin=2\tau$}.
  But the detectability of signals is independent of
  $\tau_{\rwin,\ewin}$,
  so here we keep the nominal duration fixed
  to allow for easier visual comparison.
  \label{fig:glitch_vs_psd_Q}
 }
\end{figure}

As briefly discussed before in Section \ref{sec:3.2}, 
the available energy for post-glitch tCW emission can be reduced in this model because,
regardless of the specific physical mechanisms of forming a transient mountain,
only a fraction of the superfluid excess energy is available for it.
This fraction corresponds to the healing parameter $Q$ from equation~\eqref{eq:Q},
describing the observation that most pulsars do not completely recover
their rotational frequency after a glitch.
Hence, the maximum energy radiated after a glitch via tCWs is 
\mbox{$E_\mathrm{tCW} \approx Q \Delta E_\mathrm{s}$},
which also still assumes that the glitch restores corotation.
Therefore, equation \eqref{eq:up_lim} (for rectangular signals) becomes
\begin{equation}
 \label{eq:up_lim_Q}
 h_{0,\text{r}} \leq
 \frac{1}{d}\sqrt{\frac{5G}{2c^3}\frac{\mathcal{I}}{\tau_\mathrm{r}}
 \frac{Q|\fgl|}{\frot}} \,
\end{equation}
and the factor $\sqrt{2}$ for exponential signals remains the same.
Similarly, for the ellipticity, equation~\eqref{eq:eps} becomes
\begin{equation}
 \label{eq:eps_Q}
 \epsilon_\text{r} \leq
 \frac{1}{16\pi^2} \sqrt{\frac{5c^5}{2\,G}
 \frac{Q}{\mathcal{I}\,\tau_\text{r}}\frac{|\fgl|}{\frot^5}} \,.
\end{equation}

\begin{figure}
 \includegraphics[width=\columnwidth]{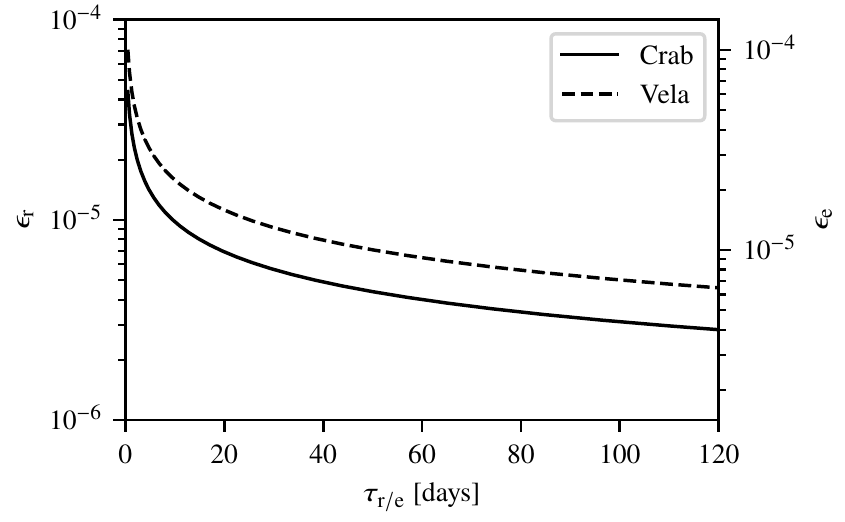}
 \caption{
  Minimum ellipticity required to detect tCWs from a transient mountain on Crab or Vela
  with a transient $\mathcal{F}$-statistic search at ET sensitivity,
  depending on the duration parameter $\tau_{\rwin/\ewin}$.
  For the left-hand vertical axis
  we assume a ``rectangular'' signal
  from a constant deformation
  ($\epsilon_\mathrm{r}$),
  while for the right-hand side
  the deformation decays exponentially
  and the plotted $\epsilon_\mathrm{e}$ is the initial value after a glitch.
  We have used approximate GW frequencies of $59.9$ and $22.4$\,Hz
  and distances of 2\,kpc and 300\,pc for Crab and Vela, respectively,
  in equation \eqref{eq:h0_eps_d}.
  \label{fig:eps_vs_tau}
 }
\end{figure}

We have used
\nglitchQATNF glitches with $Q$ listed in the ATNF glitch catalogue \citep{ATNF_glitches},
plus the Vela glitch of 2021 with parameters from \citet{Zubieta:2022umm},
to recompute upper limits.
Results are shown in Fig. \ref{fig:glitch_vs_psd_Q},
again comparing with the estimated sensitivity of transient $\mathcal{F}$-statistic searches
with the same detector configurations as before.
The number of glitches for which such searches could reach below these stricter upper limits
is significantly reduced compared with Fig. \ref{fig:glitch_vs_psd} and Table \ref{tab:1}:
with O4 sensitivity,
only \nglitchOfourYJ previous glitches from Vela and J0205$+$6449
would have been within reach,
and for O5,
\nglitchOfiveYJ glitches from 4 different pulsars
(the Crab, Vela, J0205$+$6449 and J1709$-$4429).
ET and CE could still reach below the upper limits
for similar fractions as in the generic case:
about 37 and 42 \% of glitches with known $Q$, respectively.
These two fractions, however,
are dominated by Crab and Vela glitches.

We note again, however, that inferring $Q$ from pulsar timing observations is difficult, 
as demonstrated both by the small fraction of glitches with such a value available
and by significant differences in the values listed by ATNF compared with other sources.
For example, \citet{C_D} \citep[the reference used by][]{Yim_2020}
list $Q$ values for Crab and Vela glitches until 2000.
For Vela some of these are larger than those from ATNF, 
which would correspond to more optimistic tCW upper limits.
Also, the postglitch frequency evolution model used by \citet{Yim_2020}
includes only one recovery term
(matching our equation~\eqref{eq:fgl_YJ}),
while for some glitches ATNF lists up to four pairs of $Q$ and $\tau$ parameters,
corresponding to multiple exponential recovery components fitted to that glitch.
For these cases we choose the maximum $Q$,
which ensures that the upper limit holds in either case,
and leave more detailed multi-component modelling to future work.

\subsection{Achievable reach and ellipticity constraints}
\label{sec:mountains-constraints}

Now we consider physical constraints that could be derived from tCW searches
under the transient mountain model.
We know from equation \eqref{eq:h0_eps_d} that
for any GW signal coming from a deformed NS with ellipticity $\epsilon (t)$,
and emitting at $\fgw$,
the strain amplitude $h_0 (t)$ is inversely proportional to distance;
or for a fixed distance,
it tells us the required ellipticity for a certain strain sensitivity.
To illustrate this,
in Fig. \ref{fig:eps_vs_tau}
we show the ellipticity constraints for Crab and Vela from the same equation,
assuming the sensitivity of ET,
fixing the distance and frequency to those of each pulsar,
and varying the tCW duration parameter $\tau$.
For larger $\tau$ we can detect signals from smaller asymmetries.

Since the conversion from sensitivity to constraints on distance and ellipticity
depends on the time evolution of the signal amplitude,
as before we consider both the rectangular model and the exponentially decaying case.
As discussed in Appendix \ref{sec:app_windows},
results for these two cases only differ by a factor $\sqrt{2}$,
and so in this and the following figures we again use two vertical axes for the two cases.

The obtained ellipticities, of the order of $10^{-6}$ to $10^{-4}$
for \mbox{$\tau \leq 120$\,days},
are still consistent with the latest constraints from LVK searches
for persistent CWs
from these two pulsars \citep{LIGOScientific:2021hvc},
and a transient mountain could be significantly larger still
without violating those CW upper limits.
Theoretical constraints on persistent ellipticities \citep{Johnson_McDaniel}
depend on the NS equation of state
and typically are quoted as $10^{-6}$ for conventional NS matter
and up to $10^{-5}$ for more exotic equations of state.
But for transient deformations it is conceivable that at least
the initial value could be higher.

\begin{figure}
 \includegraphics[width=\columnwidth]{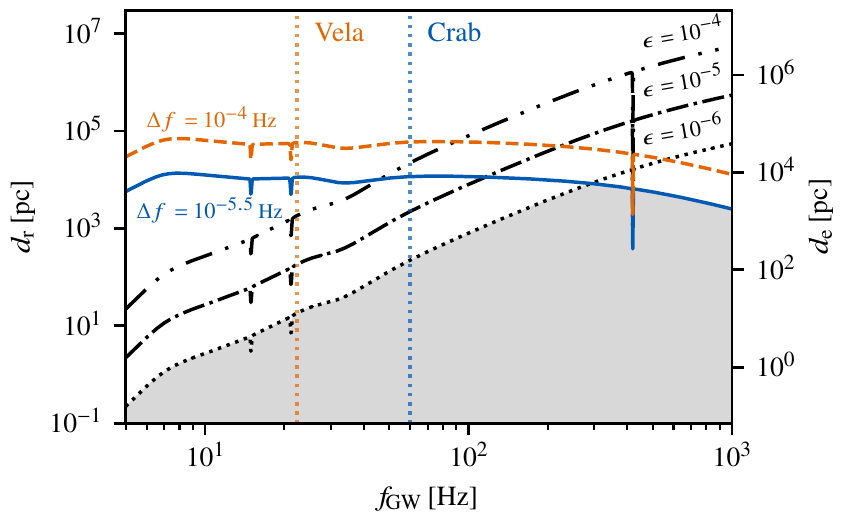}
 \caption{
  Maximum reach of a transient $\mathcal{F}$-statistic search
  with ET for tCWs from glitching pulsars,
  depending on their GW frequency
  and for an example duration parameter
  \mbox{$\tau_{\rwin/\ewin} = 10$\,days}.
  For the left-hand vertical axis
  we assume a ``rectangular'' signal,
  and an exponentially decaying one for the right-hand axis.
  We compare two different estimates here:
  The black curves
  (dotted, dash-dotted and dash-dot-dotted)
  mark the maximum distance at which we can detect a signal
  from a transient deformation with a certain (initial) ellipticity,
  from equation \eqref{eq:h0_eps_d}.
  On the other hand,
  the coloured solid and dashed curves
  indicate the maximum distance at which we can detect a signal
  with total energy corresponding to glitch sizes $\fgl$
  of $10^{-5.5}$ and $10^{-4}$\,Hz
  (corresponding approximately to the largest Crab- and Vela-like glitches, respectively),
  from equation \eqref{eq:up_lim_Q} but assuming $Q = 1$.
  For orientation,
  the frequencies of Crab and Vela are also included with dotted vertical lines.
  Detectable and physically realistic post-glitch signals
  correspond to the region under both
  a curve of fixed $\epsilon$
  and one of fixed $\fgl$.
  For example,
  the shaded area indicates the ranges for detecting signals
  with glitch size of $10^{-5.5}$\,Hz,
  $Q=1$
  and with ellipticity of $10^{-6}$.
  \label{fig:d_vs_nu}
 }
\end{figure}

\begin{figure}
 \includegraphics[width=\columnwidth]{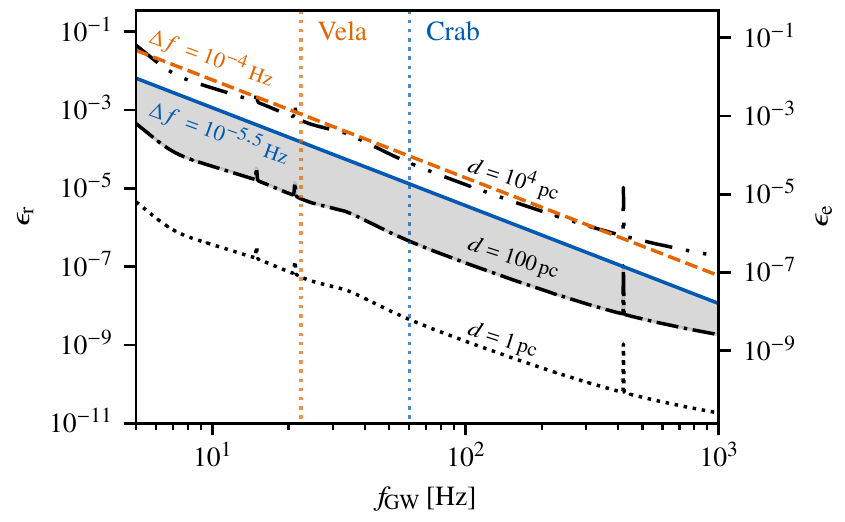}
 \caption{
  Post-glitch transient ellipticity constraints
  that could be placed if we detect a tCW of duration
  \mbox{$\tau_{\rwin/\ewin} = 10$\,days}
  from a glitching pulsar,
  depending on its GW frequency.
  For the left-hand vertical axis
  we assume a ``rectangular'' signal,
  and an exponentially decaying one for the right-hand axis.
  The black dotted, dash-dotted and dash-dot-dotted curves
  mark the minimum (initial) ellipticity
  required to detect a signal from a transient deformation
  on a pulsar at a certain distance,
  from equation \eqref{eq:h0_eps_d}
  and assuming ET sensitivity.
  On the other hand,
  the coloured solid and dashed lines
  indicate the maximum ellipticity allowed by the energy
  corresponding to glitch sizes $\fgl$ of
  $10^{-5.5}$ and $10^{-4}$\,Hz
  (corresponding approximately to the largest Crab- and Vela-like glitches, respectively)
  from equation \eqref{eq:eps_Q},
  but assuming \mbox{$Q = 1$}.
  This does not depend on detector sensitivity,
  but only on the amount of energy radiated in GWs via a transient mountain.
  For orientation,
  the frequencies of Crab and Vela themselves are also included with dotted vertical lines.
  Detectable and physically realistic post-glitch signals
  correspond to the region between
  a curve of fixed distance $d$
  and one of fixed $\fgl$.
  For example,
  the shaded area indicates the allowed ellipticities
  for detecting signals from a glitch of
  size $10^{-5.5}$\,Hz,
  $Q=1$,
  and a distance of $0.1$\,kpc.
  \label{fig:e_vs_nu}
 }
\end{figure}

While the detectability statements from Section \ref{sec:5} are independent of $\tau$
(because the detection threshold scales the same way as the available glitch energy),
Fig. \ref{fig:eps_vs_tau} gives the constraints on $\epsilon$
we could place in the case of a detection.
For this figure we have used a fit for the sensitivity depth as a function of $\tau$
(see Appendix \ref{sec:depth}).
However, this figure does not take into account whether
such ellipticities could actually be produced by a glitch of a certain size.
We will combine these considerations
in Figs. \ref{fig:d_vs_nu} and \ref{fig:e_vs_nu}.

In Fig. \ref{fig:d_vs_nu} we study the reach of tCW searches with ET, 
i.e. the maximum distance to which signals from glitching pulsars
with certain parameters can be detected,
as a function of frequency.
We compare two different estimates,
one for a signal from a transient mountain with a certain (initial) ellipticity,
from equation \eqref{eq:h0_eps_d},
and the other for a signal with total energy corresponding to certain glitch sizes,
from equation \eqref{eq:up_lim}.
(The second estimate is generic,
not relying on the transient mountain scenario.)
Both depend on the amplitude detectable by a search,
i.e. the detector noise curve scaled by the $\tau$-dependent sensitivity depth.
In addition, from equation \eqref{eq:h0_eps_d} we get a scaling with $\epsilon$ and $\fgw^2$, 
while from equation \eqref{eq:up_lim} the scaling is with $\sqrt{\fgl / \fgw}$.
If we then combine both arguments
for tCW signals emitted by a transient mountain formed after a glitch,
the joint requirement to be both detectable with ET and physically allowed by the glitch energy budget
correspond to those signals that come from a distance below both curves
for a fixed $\epsilon$ and a fixed $\fgl$.
The figure is for an example $\tau = 10$\,days.
For different durations, the curves for fixed $\epsilon$ will shift upwards
with $\sqrt{\tau}$ for larger $\tau$.
The curves for fixed $\fgl$, however, will stay the same
because the $\tau$ dependencies cancel when equating
the right-hand side of equation \eqref{eq:up_lim}
with the detectable amplitude:
both emission and detection just depend on the overall energy content,
not on its spread over time.

Similarly, in Fig. \ref{fig:e_vs_nu} we show the equivalent curves
with the ellipticity $\epsilon$ as the dependent variable.
Now the one set of estimates comes from equation \eqref{eq:h0_eps_d} with fixed distances,
and again follows the detector noise curve scaled by sensitivity depth,
but with a factor $\fgw^{-2}$.
However, the constraint on $\epsilon$ from the glitch energy budget,
as derived in equation \eqref{eq:eps},
is independent of distance and detector sensitivity,
depending only on the intrinsic parameters of the pulsar and glitch and scaling with $\fgw^{-2.5}$.
So this is a benchmark that a search needs to pass
to give physically meaningful constraints on transient mountain ellipticity. 
In other words, for a search with ET we only obtain physical constraints
in the area between two curves in Fig. \ref{fig:e_vs_nu},
namely if the ellipticity is large enough to be detectable from a given distance,
but still small enough to not exceed the available energy from a glitch.
Again, this is shown for an example
\mbox{$\tau = 10$\,days}.
Both constraints for $\epsilon$ from equations \eqref{eq:h0_eps_d} and \eqref{eq:eps}
scale with $1 / \sqrt{\tau}$, 
and so both sets of curves would shift downwards for larger $\tau$.
But consistently with Fig. \ref{fig:glitch_vs_psd},
the question of whether a certain search can reach the physical benchmark
is independent of $\tau$.
This is different from the behaviour discussed for Fig. \ref{fig:d_vs_nu}
because there the distance constraint from the glitch energy budget was generic
(not assuming the transient mountain model),
while here,
to associate an $\epsilon$ to the energy budget,
we made this mountain assumption
and that makes the $\tau$ dependence cancel in equation \eqref{eq:eps}.

In both figures, we have shown example curves for two glitch sizes 
(dashed orange: $\fgl = 10^{-5.5}$\,Hz, solid blue: $\fgl =10^{-4}$\,Hz).
We chose these as approximately the largest values for Vela- and Crab-like glitches, respectively.
For these figures we have implicitly used a factor $Q=1$.
According to \citet{ATNF_glitches},
Crab glitches have typical \mbox{$Q \approx 0.8$},
but typical Vela glitches have much lower values,
at most \mbox{$Q \sim 0.1$}.
Thus, for realistic Vela glitches the curves in both figures
could actually fall below those for Crab glitches.

This pair of plots
(Figs. \ref{fig:d_vs_nu} and \ref{fig:e_vs_nu})
corresponds to the typical plots of
$d (\fgw)$ at fixed $\epsilon$
and $\epsilon (\fgw)$ at fixed $d$
used in CW papers \citep[e.g.][]{O3-2_Abbott_2021, LIGOScientific:2022pjk}.
But as we have seen, for tCWs from glitching pulsars
this type of plot depends on additional physical parameters,
i.e. the glitch size and signal duration.

\subsection{Model checks against observed glitch population}
\label{sec:mountains-checks}

We now discuss the \citet{Yim_2020} model assumptions in more detail
and investigate how well the model matches the available pulsar glitch data.

When we introduced equation \eqref{eq:fgl_YJ} for the frequency change at a glitch,
following \citet{Yim_2020},
we included both
a transient $\Delta f_{\text{gl,p}}$
and permanent part $\Delta f_{\text{gl,t}}$
but only a permanent part $\Delta \dot{f}_{\text{gl,p}}$ for the spindown change.
Taking its derivative with respect to time one obtains the equation for
the total spindown change $\fdotgl$ for times
\mbox{$t\geq t_\text{g}$}:
\begin{equation}
 \label{eq:fdotgl_YJ}
 \fdotgl(t) = \Delta \dot{f}_{\text{gl,p}}
 - \frac{\Delta f_{\text{gl,t}}}{\tau}\,e^{-\frac{\Delta t}{\tau}} \,.
\end{equation}
From this we can rewrite the time-dependent coefficient as the transient change in spindown,
\begin{equation}
 \label{eq:fdot_trans} 
 \Delta \dot{f}_{\text{gl,t}} = -\frac{\Delta f_{\text{gl,t}}}{\tau} \,.
\end{equation}

Rearranging this equation and taking also equation \eqref{eq:Q} one can deduce 
\begin{equation}
 \label{eq:Q_tau}
 \Delta \dot{f}_{\text{gl,t}}\,\tau = - \Delta f_{\text{gl,t}} = -Q \fgl\,.
\end{equation}
This relates the transient spindown change with 
observationally more accessible quantities:
the factor $Q$, the total frequency change of the pulsar at the glitch,
and the recovery timescale $\tau$.

The central assumption of \citet{Yim_2020} is to approximate 
\mbox{$\Delta \dot{f}_{\text{gl,t}} \approx \Delta \dot{f}_\text{gl}$}. 
That is, to assume that the change in spindown is purely transient, 
without any permanent part.
Under this assumption,
they define an approximate timescale
\begin{equation}
 \label{tau_approx}
 \tau_\text{approx} = -\frac{Q\fgl}{\fdotgl} = 
 \frac{\Delta\dot{f}_{\text{gl,t}}}{\fdotgl}  \tau\,.
\end{equation} 
This again only depends on parameters that can be directly inferred from pulsar timing.
Physically, the approximation corresponds to assuming that no new permanent mountain 
is formed
(or an existing one permanently augmented)
at the time of the glitch.
In other words, any non-axisymmetry of the pulsar corresponds only 
to a truly persistent part independent of the glitch
(which would emit traditional CWs) 
and a transient part created by the glitch, 
that decays again during the recovery timescale.

Comparing $\tau_{\text{approx}}$ against $\tau$ for observed glitches
can be used as a consistency check of the model.
Even if they do not agree within uncertainties,
the case \mbox{$\tau_{\text{approx}} < \tau$}
corresponds to
part of the formed mountain being permanent,
i.e. events for which one can not fully apply this approximation, 
but which are still physically reasonable.
On the other hand,
the case \mbox{$\tau_{\text{approx}} > \tau$}
is physically inconsistent with the transient mountain scenario.

In \citet{Yim_2020},
this approximation has been tested for the Crab and Vela pulsars,
using benchmarks as discussed below.
They obtained generally consistent results,
except for the larger Crab glitches and some Vela glitches.
Large glitches of the Crab are the only known pulsar glitches
that exhibit a measurable permanent change in spindown \citep{Lyne_93abc, Lyne:2000sta}.
Meanwhile many Vela glitch recoveries require to be modelled by more than one exponential,
each with its own $\tau$,
and some of its glitches overlap with each other,
complicating the estimation of $Q$.

Here we extend the comparison of their model with observational data
to the larger set of \nglitchQATNF glitches
with recovery parameters ($\tau$ and $Q$)
listed in the ATNF pulsar glitch catalogue,
plus the 2021 Vela glitch from \citet{Zubieta:2022umm}.
(These parameters are not included in the Jodrell Bank glitch catalogue.)
To find how well the assumptions of the model match each glitch, 
we have first compared the tabulated values of $\tau$
(including uncertainties)
with $\tau_\text{approx}$ calculated from equation \eqref{tau_approx}
(including error propagation).
As in Section~\ref{sec:mountains-uls},
for glitches with multiple recovery components
we have chosen the one with the highest $Q$
and the corresponding $\tau$.
We find consistent values for 59 glitches out of \nglitchQ,
in the sense that the two uncertainty ranges overlap.
Moreover, from the 60 cases with inconsistent $\tau$, 
47 have \mbox{$\tau_\text{approx} < \tau$},
i.e. are still consistent with a combination of
transient and permanent mountain formation.
In other words, from 
\nglitchQ glitches, 
only 13 show \mbox{$\tau_\text{approx} > \tau$},
being inconsistent with the transient mountain scenario.

Following this, another consistency check for the model
is whether the transient mountains assumed to be formed by a glitch
are physically possible, 
i.e. that the model predicts realistic values of the initial ellipticity
right after the glitch.
From equation~\eqref{eq:eps_Q}, 
using equation \eqref{eq:Q_tau}
the $\sqrt{Q}$ factor cancels:
\begin{equation}
 \label{eq:eps_YJ}
 \epsilon_\text{approx} = 
 \sqrt{\frac{- 5}{2^{9}\pi^4}\frac{c^5}{G\mathcal{I}}\frac{\fdotgl}{\frot^5}} \,.
\end{equation}

With this we can calculate the expected ellipticity of the same \nglitchQ glitches.
We can then revisit the $\tau$ check excluding glitches with unrealistically large ellipticities,
which can happen especially for low $\frot$.
As discussed in Section \ref{sec:mountains-constraints},
transient mountains could conceivably be larger
than the usual constraints on persistent mountains.
Also if we impose too strict a cut then the number of selected glitches
would be too small to make useful statements.
Therefore we choose a rather high example cutoff of
\mbox{$\epsilon < 0.001$}.
This leaves \nglitchQloweps glitches with available recovery parameters and low $\epsilon$,
from which 37 have consistent values of $\tau_\text{approx}$ and $\tau$,
and a further 31 have
\mbox{$\tau_\text{approx} < \tau$}.
Hence the fraction of physically reasonable cases only slightly increases
after the cut on $\epsilon$:
68 out of \nglitchQloweps glitches.

Finally, we can check another of the assumptions of \citet{Yim_2020}, 
namely that the change in the angular momentum at the glitch
is only due to a change in spin frequency,
neglecting any change in the pulsar's moment of inertia.
The variables $\eta_1$ and $\eta_2$ defined in equations (B16) and (B17)
of \citet{Yim_2020}\footnote{
There is a factor of $1/4$ missing on the right hand sides of their two equations,
as confirmed by the authors 
(private communication).
As written here,
our equations \eqref{eq:eta1} and \eqref{eq:eta2} are consistent with
the left hand sides of equations (B16) and (B17) of \citet{Yim_2020},
as well as their equation (B13).}
tell us how relevant the changes in the moment of inertia are
in the contribution to the angular momentum at the glitch
and during recovery after the glitch,
respectively:
\begin{equation}
 \eta_1 =
 \frac{- 5}{2^{11}\pi^4}\frac{c^5}{G\mathcal{I}}\frac{\fdotgl}{\frot^4\fgl} \,;
 \label{eq:eta1}
\end{equation}
\begin{equation}
 \eta_2 = 
 \frac{- 5}{2^{11}\pi^4}\frac{c^5}{G\mathcal{I}}\frac{1}{Q}\frac{\fdotgl}{\frot^4\fgl} \,.
 \label{eq:eta2}
\end{equation}

As suggested by \citet{Yim_2020},
values of \mbox{$\eta_{1,2} < 1$}
would imply that
neglecting the corresponding changes in the moment of inertia is safe, 
while larger values indicate
that they would contribute significantly to the pulsar's spin-down.
Considering again only the \nglitchQloweps glitches
with known recovery parameters and with low $\epsilon$,
we find that all have \mbox{$\eta_1 < 1$}
and 58 also have \mbox{$\eta_2 < 1$}.
On the other hand, if we do the same test with all 
\nglitchQ glitches with recovery parameters, we find
101 glitches with \mbox{$\eta_1 < 1$},
64 of which also have \mbox{$\eta_2 < 1$}.
In contrast with large values of $\epsilon$,
\citet{Yim_2020} suggest that large values of these parameters
do not necessarily indicate an unrealistic scenario, 
but can be compensated by a larger initial event causing the glitch.
Hence, one can still consider the transient mountain emission scenario for those glitches,
but should be careful with the detailed physical interpretation
of putative tCW signals from them.

As a side note, for Crab and Vela,
we have obtained some numerical results different from theirs.
These are mainly due to different values of $Q$
listed in our source
(the ATNF glitch catalogue)
compared to their source \citep{C_D}.

Summarizing, the model by \citet{Yim_2020}
is unlikely to apply to the entire glitch population, 
but it is consistent with a large fraction of the glitches
with reported recovery parameters.
A more systematic study could investigate how deviations from the model
correlate with various pulsar and glitch parameters, 
and how these discrepancies should be interpreted physically.

\section{Conclusions}
\label{sec:conc}

We have estimated detection prospects
for quasi-monochromatic long-duration GW transients (tCWs),
based on the overall energy budget of pulsar glitches
and realistic tCW search setups with the
transient $\mathcal{F}$-statistic method~\citep{Prix_G_M, Keitel_2019, our_paper, Luana2021}.
We have used a combined set of $\nglitchusable$ glitches,
mainly from the ATNF and Jodrell Bank catalogues~\citep{ATNF_glitches, Jodrell_glitches,Espinoza_2011,Jodrell_extra}.

General prospects are summarised in Fig.~\ref{fig:glitch_vs_psd}
under the optimistic assumption of using the whole superfluid excess energy from before a glitch
as the budget for post-glitch tCW emission,
independent of the tCW emission mechanism.
We have found promising prospects for detecting tCWs
with the next observing runs of the current LVK detector network or with next-generation detectors,
or at least for reaching a physically constraining regime
where observational upper limits are below these generic indirect limits.
In the O4 and O5 runs of the current LVK detector network,
approximately 5\% and 10\% of known past glitches could be in reach
if those pulsars glitch again with similar parameters.
Moreover, for the third-generation detectors ET and CE,
this fraction increases up to 35--40\%.

But we have also explored the more specific assumption
of a transient mountain as the emission mechanism,
following the model introduced by \citet{Yim_2020}.
On the one hand, this leads to stricter indirect upper limits,
as only a fraction of the energy budget is available for tCW emission from the transient mountain,
determined by the glitch healing parameter \mbox{$0\leq Q \leq1$}.
Together with the smaller number of \nglitchQ glitches for which this parameter is available,
only \nglitchOfourYJ and \nglitchOfiveYJ previously observed glitches
would be accessible in O4 and O5, respectively,
while with ET and CE still about 40\% of those with known $Q$
could be within reach.
On the other hand, this model allows studying
the constraints on physical source parameters that could be obtained
by requiring both
(i) detectability with a given detector and search setup
and (ii) maintaining the glitch energy budget.
This yields regions of detectable and physically realistic tCW signals
in the space of signal frequency, ellipticity and distance,
as illustrated in Figs. \ref{fig:d_vs_nu} and \ref{fig:e_vs_nu}.

We emphasize that any such estimation of detection prospects depends on
some choice of threshold for declaring a signal as detectable.
Here, we estimated the detection threshold using simulated noise distributions
for realistic narrowband transient-$\mathcal{F}$-statistic search setups.
This approach takes the ``trials factor'' of a search properly into account~\citep{Tenorio:2021wad}.
We recommend using a similar approach whenever possible,
rather than referring to a generic signal-to-noise ratio threshold,
to ensure that
predictions realistically represent actual search capabilities.

We have also extended tests of the transient mountain model's physical consistency,
which \citet{Yim_2020} previously applied to Crab and Vela glitches,
to all glitches with recovery parameters included in the ATNF catalogue.
We have found that approximately half of the analysed glitches
are fully consistent with the model assumption of a purely transient change in spindown
(corresponding to a purely transient mountain formed at the glitch).
Most others are at least consistent with a mix
of transient and permanent contributions,
still allowing for some (reduced) tCW emission under this model.

A detailed investigation of the model-specific upper limits
for other tCW emission mechanisms,
such as Ekman flows or $r$-modes
(see works cited in \ref{sec:tCWs_basics} and references therein),
is beyond the scope of this work.
The generic upper limits would still be expected
to cover those cases too,
but it would be interesting to compare them more quantitatively
against the transient mountain scenario.

On the observational side,
searches for tCWs have so far been limited to narrowband analyses
of known pulsars~\citep{Keitel_2019, our_paper}.
Next-generation pulsar surveys, e.g.
with FAST~\citep{FAST},
CHIME~\citep{CHIME}
and the SKA~\citep{SKA}
should enlarge the set of targets for such sources.
But other search types over wider parameter spaces could be considered,
with the most ambitious goal of blind all-frequency all-sky searches.
Besides being less sensitive due to the much higher trials factor,
a currently prohibitive issue for such broad searches with
the transient $\mathcal{F}$-statistic method is computational cost \citep{Keitel_Ashton_2018}.
Alternative semi-coherent or machine-learning methods
\cite[e.g.][]{Keitel_2015,Suvorova:2016rdc,2019PhRvD.100b3006B,Cuoco,Modafferi_CNNs}
could make them accessible,
but likely still at lower sensitivities than
the narrowband coherent matched-filter searches considered here,
which hence serve as an overall reference benchmark for the field.

\section*{Acknowledgements}

We thank Garvin Yim and Ian Jones for detailed discussions
of their model
and on ways to plot sensitivity estimates,
Benjamin Shaw for help in
tracking down pulsar aliase
s and rapid corrections of minor typos in the Jodrell Bank glitch catalogue,
and the CW working group of the LVK collaborations for discussions and feedback.
This work has been supported by European Union FEDER funds;
the Spanish Ministry of Science and Innovation
and the Spanish Agencia Estatal de Investigaci{\'o}n Grants No.
PID2019–106416GB-I00/MCIN/AEI/10.13039/501100011033,
RED2018-102661-T,
RED2018-102573-E;
the European Union NextGenerationEU funds (PRTR-C17.I1);
the Comunitat Aut{\`o}noma de les Illes Balears
through the Direcci{\'o} General de Pol{\'i}tica Universitària i Recerca
with funds from the Tourist Stay Tax Law ITS 2017-006
(PRD2018/24, PDR2020/11);
the Conselleria de Fons Europeus, Universitat i Cultura del Govern de les Illes Balears;
the Generalitat Valenciana (PROMETEO/2019/071);
and EU COST Actions CA18108 and CA17137.
DK is supported by the Spanish Ministerio de Ciencia, Innovaci{\'o}n y Universidades
(ref. BEAGAL 18/00148)
and cofinanced by the Universitat de les Illes Balears.
RT is supported by the Spanish Ministerio de Ciencia, Innovaci{\'o}n y Universidades
(ref. FPU 18/00694). 

The authors gratefully acknowledge computational resources provided by
the LIGO Laboratory
and supported by
National Science Foundation Grants PHY-0757058 and PHY-0823459,
and at Artemisa,
funded by the European Union ERDF and Comunitat Valenciana,
as well as the technical support
provided by the Instituto de Física Corpuscular, IFIC (CSIC-UV).

This paper has been assigned document numbers
\href{https://dcc.ligo.org/\dcc}{LIGO-\dcc}
and \href{\tdsurl}{\tds}.

\section*{Data Availability}

All primary data used in this article were taken from the public
ATNF pulsar catalogue~\citep{Manchester:2004bp},
ATNF glitch catalogue~\citep{ATNF_glitches},
and Jodrell Bank glitch catalogue~\citep{Jodrell_glitches,Espinoza_2011,Jodrell_extra};
as well as results from \citet{UTMOST_2} and \citet{Zubieta:2022umm}.
Results shown in our figures and tables
can be easily reproduced from those inputs
with the equations in this paper,
but tabular data can also be provided upon request.


\bibliographystyle{mnras}
\bibliography{prospects}



\appendix

\section{Processing of glitch catalogues}
\label{sec:A0}

The ATNF~\citep{ATNF_glitches} and Jodrell~\citep{Jodrell_glitches,Espinoza_2011,Jodrell_extra} glitch tables
have substantial overlap,
though both tables also contain unique entries.
To get the most representative view of the currently known glitch population,
we merged the two tables while removing duplicate entries.
We also match glitches to pulsars in the main ATNF pulsar catalogue~\citep{Manchester:2004bp}
by name to obtain parameters such as distance and frequency.

We find that all the events in the ATNF glitch catalogue have a match in the pulsar catalogue,
and for \nglitchJodrellmatched out of the \nglitchJodrell glitches in the Jodrell catalogue
we also find correspondence.
The remaining entries
for which we were not able to match them with pulsar entries in the ATNF catalogue
even after some degree of manual checks for spelling variations are
the extragalactic X-ray pulsar M82-X2~\citep{2020ApJ...891...44B}
and the anomalous X-ray pulsar J1048$-$5937~\citep{Dib:2008ts}.
These cases are excluded from our analysis.

Then, to remove duplicates between the two glitch catalogues,
we used the ATNF list as the default,
i.e. the parameters listed in it take preference for glitches listed in both tables.
We considered two entries from the two catalogues with matching pulsar names as the same glitch
if their listed glitch epochs have less than 1 day of difference,
or if their uncertainty intervals overlap.

To make sure that we are not missing any repeated glitches,
we further manually analysed the 18 pairs of glitches with less than 10 days of difference
that have not been removed with the previous step.
Comparing the original references for each,
we found that 17 of these can be considered as the same.
The remaining one from J2301+5852
presents considerably different values of the relative glitch sizes,
but even so the dates are very similar --
only slightly more than one day of difference.
Hence we decided to consider this one as a duplicate as well,
falling back to the ATNF parameters as usual.

We are left with the \nglitchATNF ATNF glitches
and \nglitchJodrellunique additional Jodrell glitches.
We also added another glitch of J1105$-$6910
observed by the UTMOST pulsar timing programme \citep{UTMOST_2},
which was also analysed in \citet{our_paper},
and the recent July 22nd 2021 Vela glitch
analysed in \citet{Zubieta:2022umm}.
This gives us \nglitchtotal glitches in total
from \npulsarstotal pulsars.

Among these, however,
there are some cases with missing parameters that we need for our analyses.
For two glitches from J1341$-$6220 there are no individual glitch size measurements,
as discussed in \citet{Lower:2021rdo}.
Two events from J2301+5852 are ``anti-glitches''
with \mbox{$\fgl < 0$},
for which our analysis is not appropriate.
Ten more glitches are from 8 pulsars without distances listed in the ATNF pulsar catalogue.

In the end, we use \nglitchusable glitches
with sufficient information for our main analysis
in Section \ref{sec:5},
coming from \npulsarsusable pulsars.

\vspace{-\baselineskip}

\section{Subpopulations of glitches}
\label{sec:A2}

As discussed in Section~\ref{sec:obs_pop},
the observed glitch population is often categorised into two subpopulations
of smaller ``Crab-like'' and larger ``Vela-like'' glitches,
see e.g.~\citet{main_2}.
To give an updated overview of the latest combined data set of pulsar glitches from the
ATNF and Jodrell Bank glitch catalogues~\citep{ATNF_glitches, Jodrell_glitches,Espinoza_2011,Jodrell_extra}
as we use them in this work,
we provide here two standard plots of relative and absolute glitch sizes
in frequency and spindown.
Here we use a smaller set (\nglitchspd glitches from \npulsarsspd pulsars)
than in our main analyses because spindown changes are not catalogued for all events.
The transition value of \mbox{$\fgl = 10^{-5.5}$\,Hz} from~\citet{main_2}
between the two subpopulations
(obtained from a Gaussian kernel density estimation fit)
is re-used for illustrative purposes,
without redoing such a fit.

\begin{figure}
 \includegraphics[width=\columnwidth]{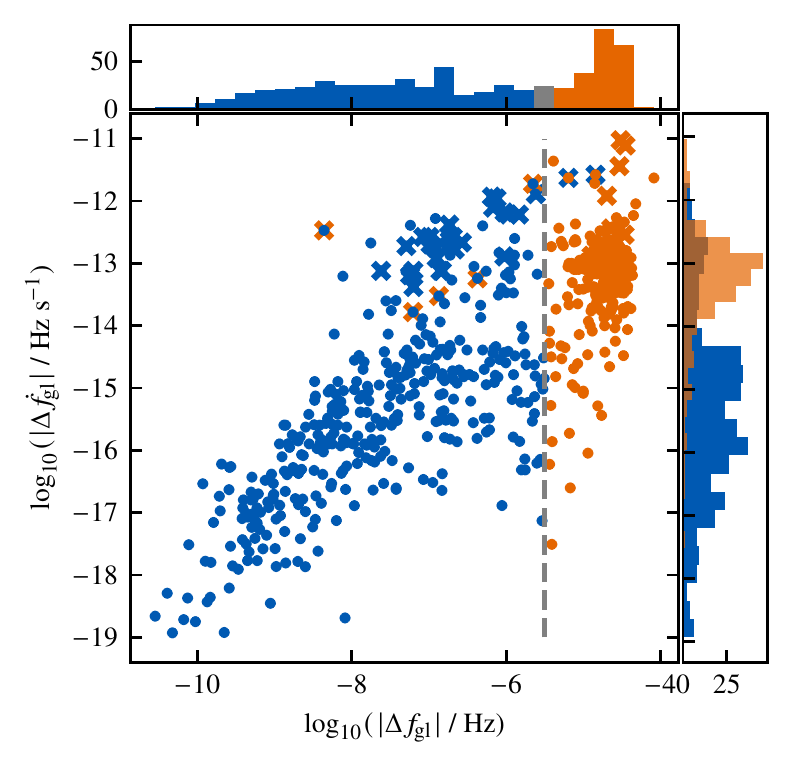}
 \caption{
  Pairs of $\fdotgl$ and $\fgl$
  together with histograms of both variables
  for \nglitchspd known pulsar glitches.
  In blue, Crab-like glitches,
  and in orange, Vela-like glitches,
  according to the transition scale
  \mbox{$\fgl = 10^{-5.5}$\,Hz} from \citet{main_2},
  indicated by the dashed line.
  \label{fig:dsd_vs_dfreq}
 }
\end{figure}

First, Fig.~\ref{fig:dsd_vs_dfreq} shows the distribution of absolute glitch sizes,
where one can see not only the relatively clear two-Gaussian distribution of frequency steps,
but also the correlation between this variable and the spindown change.
Vela-like glitches tend to present bigger values of $\fdotgl$
in a narrower distribution than Crab-like glitches,
which also tend to present smaller values of spindown change.
On the other hand, there is no clear relation between frequency step size
and the original spindown of the source,
as was also concluded in \citet{main_2}. 
Similarly, the spindown change is also not clearly correlated
with the original spindown of the pulsar.

\begin{figure}
 \includegraphics[width=\columnwidth]{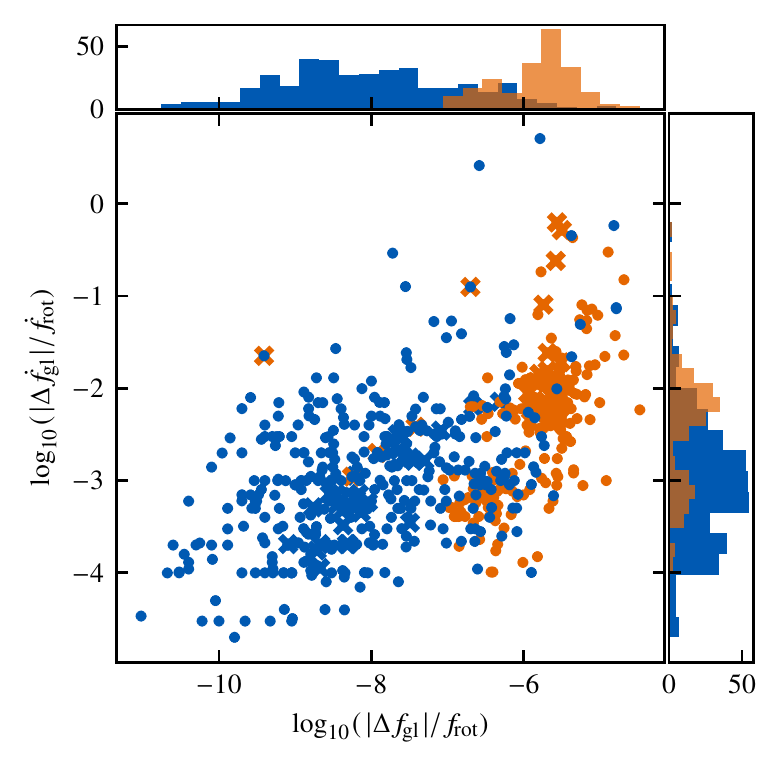}
 \caption{
  Pairs of $\fdotgl/\fdotrot$ and $\fgl/\frot$
  together with histograms of both variables
  for \nglitchspd known pulsar glitches.
  In blue, Crab-like glitches,
  and in orange, Vela-like glitches,
  according to the transition scale
  \mbox{$\fgl = 10^{-5.5}$\,Hz} from \citet{main_2}.
  \label{fig:dsd_vs_dfreq_rel}
 }
\end{figure}

In Fig.~\ref{fig:dsd_vs_dfreq_rel} we plot the glitch step sizes variables
divided by their original value,
i.e. the relative glitch sizes in frequency and spindown,
applying the same categorisation from \citet{main_2}
as in Fig. \ref{fig:dsd_vs_dfreq}.
Both the subpopulation distinction and the correlation are less pronounced
in these variables.

As already discussed in Section~\ref{sec:obs_pop},
we do not make any explicit distinctions in this paper
based on the $\fgl$ separation criterion.
Rather, we just treat Crab and Vela as prototypes
for pulsars with smaller and larger glitches, respectively, 
but taking into account that not all glitches from either pulsar fall
on the same side of that division line.

For other studies of pulsar glitch population statistics,
see e.g. \citet{Espinoza_2011,main_1,Lower:2021rdo,Jodrell_extra,Arumugam:2022ugq}.

\vspace{-\baselineskip}

\section{Energy budget for post-glitch tCW emission}
\label{sec:energy_budget}

Here we explain in more detail the use of $\Delta E_\textrm{s}$
from equation~\eqref{eq:E_super}
as the available energy budget for post-glitch tCWs. 
We use observational constraints on the glitch size,
assume that a glitch restores the corotation between the normal 
and superfluid components of the NS, and
assume the typical relation between fiducial
moments of inertia as previously done by~\citet{Prix_G_M}.

Following the two-fluid model as described in \citet{Prix_G_M},
a glitch corresponds to 
an energy transfer between the superfluid component and the normal component
of a neutron star: the superfluid component slows down by $\delta \Oms$
(a negative number)
while the normal component spins up by
\mbox{$\delta \Omega = 2 \pi \fgl$}.

Typically observed spin-up values of the normal component
are such that at most
\mbox{$\delta \Omega / \Omega \sim 10^{-6}$},
while no direct observational constraints
are available for $\Delta \Omega$ or $\delta \Oms$.
Under the standard assumption on
the fiducial values of the moments of inertia
\mbox{$\mathcal{I}_\textrm{s} / \mathcal{I} \sim 10^{-2}$},
conservation of angular momentum implies
\mbox{$-\delta \Oms / \Omega 
\approx (\mathcal{I} / \mathcal{I}_{\textrm{s}}) \delta \Omega /\Omega \sim 10^{-4}$}.
This implies $\delta \Omega_{(\textrm{s})}$ can both be neglected to leading order
so that e.g. \mbox{$\Omega_\textrm{s} + \Omega \approx 2 \Omega$}.
The assumption that corotation is restored after the glitch implies
\mbox{$\Delta \Omega = - \delta \Omega_{s} + \delta \Omega$},
which allows us to relate lag and glitch size as
\mbox{$\Delta \Omega = \delta \Omega (1 - \delta \Oms /\delta \Omega) 
\approx \delta \Omega \, \mathcal{I} / \mathcal{I}_\textrm{s}$}.

The energy required to decrease the rotational energy of 
the superfluid component by $\delta \Omega_{\textrm{s}}$ is
\begin{equation}
    \label{eq:E_co}
    \Delta E_{\textrm{co}} 
    = \frac{1}{2}\mathcal{I}_{\textrm{s}}[
        \Omega_{\textrm{s}}^2 - (\Omega_{\textrm{s}}
        + \delta\Omega_{\textrm{s}})^2]
    = -\frac{1}{2}\mathcal{I}_\textrm{s}(
    2 \Omega_\textrm{s} \delta \Omega_{\textrm{s}} 
    + \delta \Omega_{\textrm{s}}^2)\;,
\end{equation}
which agrees to leading order with the energy required
to spin up the normal component
by $\delta \Omega$ if corotation is assumed to be restored.
By comparing equations \eqref{eq:E_super} and \eqref{eq:E_co} we obtain
\begin{equation}
    \frac{\Delta E_{\textrm{co}}}{\Delta E_{\textrm{s}}}
    = \frac{-2 \Oms \delta \Oms - \delta \Oms^2}
    {(\Oms - \Omega)(\Oms + \Omega)}
    \approx -\frac{\delta \Omega_{s}}{\Delta \Omega}
    \approx 1 + \frac{\delta \Omega}{\delta \Oms}\;.
\end{equation}
Angular momentum conservation imposes 
\mbox{$\delta \Omega / \delta \Oms = 
-\mathcal{I}_\textrm{s} / \mathcal{I} \sim -10^{-2}$}
which implies most of the superfluid excess energy must be transferred directly 
into the normal component during a glitch.
Consequently, we take $\Delta E_\textrm{s}$ from equation~\eqref{eq:E_super}
as the relevant budget
to place upper bounds on post-glitch tCW emission.
Dropping the assumption of restoring corotation after each glitch
would be equivalent to reducing the available budget 
$\Delta E_{\textrm{s}}$, 
with the specific fraction being highly dependent
on the specifics of the glitch model
and thus left for future work. 

Comparing to the derivation in \citet{Prix_G_M},
their $E_{\textrm{s}}$ corresponds to our $\Delta E_{\textrm{s}}$, 
while their $E_{\textrm{glitch}}$
would correspond to $\Delta E_{\textrm{s}} - \Delta E_{\textrm{co}}$
and is the relevant budget for short-duration GW emission
directly triggered by the glitch
\citep{Ho:2020nhi, Lopez:2022yph}.

\vspace{-\baselineskip}

\section{Rectangular vs. exponential tCWs}
\label{sec:app_windows}

Here we clarify some of the relations between tCW signals
with either a rectangular window (constant amplitude) or exponential window,
following the conventions introduced by \citet{Prix_G_M}.

We start from equation \eqref{eq:E_gw} for the total energy in a GW signal,
which is proportional to the integral of the squared signal amplitude over the signal duration,
and hence also proportional to the square of the root-mean-square amplitude
from equation \eqref{hhat_def}:
\begin{equation}
 \label{18}
 \Egw \propto \hat{h}^2_0 \,\, T = \int_0^T h_0^2(t) \, \mathrm{d}t \,.
\end{equation}
Analogously to equation \eqref{eq:signalmodel} for the full $h(t)$,
through $h_0(t)$ this also depends on the tCW window function.
For a rectangular window,
the integral goes to \mbox{$T=\tau_\rwin$},
which by definition represents the total duration of the signal.
In this case, the integral becomes trivial:
\begin{equation}
 \label{19}
 \Egw \propto \int^{\tau_\rwin}_0 h_{0,\rwin}^2 \,\, \mathrm{d}t = h^2_{0,\rwin} \tau_\rwin \,.
\end{equation}

For an exponential window,
following the convention from \citet{Prix_G_M}
to cut it after \mbox{$T=3\tau_\ewin$},
the same quantity becomes
\begin{equation}
 \label{20}
 \Egw \propto \int^{3\tau_\ewin}_0 h_{0,\ewin}^2 e^{\frac{-2t}{\tau_\ewin}}\,\, \mathrm{d}t
 = h^2_{0,\ewin} (1-e^{-6}) \frac{\tau_\ewin}{2} \approx h^2_{0,\ewin} \frac{\tau_\ewin}{2} \,.
\end{equation}
(The exact choice of the cutoff does not matter as long as $\exp(-2T/\tau_\ewin)\ll1$.)

\begin{figure}
  \includegraphics[width=\columnwidth]{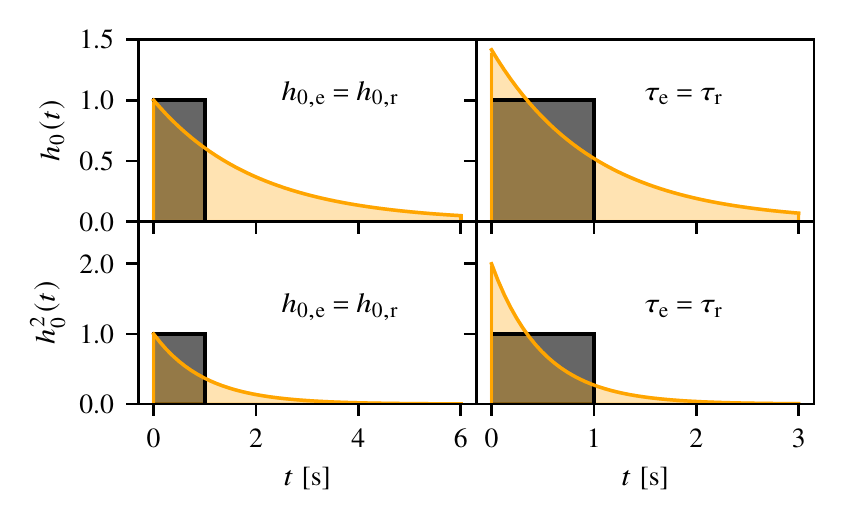}
  \caption{Comparison between exponential and rectangular window functions.
   In the first column,
   the two signals with different windows
   have the same initial strain amplitude $h_{0,\rwin/\ewin}$,
   and duration parameters $\tau_{\rwin/\ewin}$ are computed
   from equations~\eqref{19} and \eqref{20}
   to ensure equal total energy.
   In the second column,
   $\tau_{\rwin/\ewin}$ are equal for the two windows,
   and $h_{0,\rwin/\ewin}$ are computed
   from the same equations for equal total energy.
   The top row shows the strain amplitude $h_0(t)$,
   where areas under the curves are \emph{not} equal
   between the two windows,
   while in the bottom area $h_0^2(t)$ is shown,
   with the areas under the curves equal
   and proportional to the total GW energy.
   \label{fig:windows}
  }
\end{figure}

Now we will consider signals with either rectangular and exponential shape
emitted by the same source, i.e. with the same total energy,
and compare the two windows in two special cases
which are also illustrated in Fig.~\ref{fig:windows}:
either equal duration parameters $\tau$
or equal initial amplitudes $h_0$.
For \mbox{$\tau_\ewin = \tau_\rwin$},
we find that equations~\eqref{19} and \eqref{20} enforce
$h_{0,\ewin} =  \sqrt{2} h_{0,\rwin}$
to have the same total energy in both signals;
while for \mbox{$h_{0,\ewin} = h_{0,\rwin}$}
the requirement is
$\tau_\ewin = 2 \tau_\rwin \,.$
This can be visually appreciated in the bottom row of Fig.~\ref{fig:windows},
where areas beneath the curves correspond to the energy,
and are equal for the two windows.

We also verified empirically that these relations hold
for the implementation of the $\mathcal{F}$-statistic
in \texttt{LALSuite} \citep{LALSuite}
using the \texttt{synthesizeTransientStats} program \citep{Prix_G_M}.
We simulated signals at different initial $h_0$
and a fixed duration parameter \mbox{$\tau_{\rwin/\ewin}=10$\,days}
with the two windows.
We find the value of $h_{0,\rwin/\ewin}$ for which 95\% of the signals
fall above a certain detection statistic threshold.
(We used the same threshold as derived for a realistic simulated search
below in Appendix \ref{sec:depth}.)
The initial test produced a ratio of
\mbox{$h_{0,\ewin}^{95\%} / h_{0,\rwin}^{95\%}\approx1.28$},
instead of the expected \mbox{$\sqrt{2} \approx 1.4$}.
The reason was in the timestamps we were using for the simulated analysis,
from the O3 run.
Their gaps (especially the month-long maintenance break in October 2019)
mean that at equal $\tau$ an exponential window
(with a total duration cut at \mbox{$T=3\tau_\ewin$})
does not actually accumulate as much additional signal power
compared to the rectangular window with duration \mbox{$T=\tau$}
as in ideal uninterrupted data.
With gapless data, the empirical ratio instead is found as $\approx 1.4$.

The main conclusion from the comparison in this Appendix is that
our main results, especially the indirect energy upper limits
as in Fig. \ref{fig:glitch_vs_psd},
can be converted from rectangular to exponential windows
by rescaling any $h_{0,\rwin}$ values
with a factor $\sqrt{2}$,
as done in the right-hand side vertical axis of that figure.
The same scaling applies to the inferred ellipticity constraints
in Figs.~\ref{fig:eps_vs_tau} and \ref{fig:e_vs_nu},
and the inverse scaling applies to the sensitivity depth $\mathcal{D}$
from equation~\eqref{eq:depth}
and the distance constraints in Fig.~\ref{fig:d_vs_nu}.

\vspace{-\baselineskip}

\section{Sensitivity depth estimation}
\label{sec:depth}

To properly determine the expected sensitivity curves of 
realistic tCW searches with future detectors 
and compare them with indirect energy upper limits of known pulsar glitches,
we have to estimate the sensitivity depth $\mathcal{D}$ defined in 
equation~\eqref{eq:depth}.
We do this for two prototypical post-glitch searches,
namely those targeting the Crab and Vela pulsars.

Candidates from searches over a template bank,
such as the transient-$\mathcal{F}$-statistic searches we consider here,
are usually identified by setting a threshold at a fixed false-alarm probability
on the \emph{overall} search, which results in a higher value than that corresponding
to a single-template false-alarm probability.
Such a threshold, which depends on the extension of the searched parameter space
(hence on the number of templates analysed by a search), can be easily estimated
from actual or simulated search results
after applying \texttt{distromax}~\citep{Tenorio:2021wad, distromax}.
With this information, and using a realistic duty factor for the detectors,
we can use the \citet{Dreissigacker2018} method
to estimate the depths $\mathcal{D}$ for different detector configurations.

\vspace{-\baselineskip}

\subsection{Number of templates}
\label{sec:depth-templates}
We consider the Vela glitch of 2016 December 12 \citep{Palfreyman:2018gli}
previously analysed by \citet{Keitel_2019}
and the Crab glitch of 2019 July 23 \citep{Shaw:2021vvs}
analysed by \citet{our_paper}.
Those were narrowband searches allowing for some flexibility of GW frequency evolution
around the pulsar timing solutions,
i.e. \mbox{$\fgw\approx2\frot$} and similarly for the spindown parameters.
As in those analyses, we assume a search of 120 days of data after the glitch.
Following the setup in \citet{Luana2021},
the widths of frequency and spindown bands suitable for each search
are calculated from the pulsar ephemerides.
For the Crab the result is the same as in \citet{our_paper},
where the procedure from \citet{Luana2021} was already used:
\mbox{$\Ntemp \approx 3.9\times10^{6}$} templates.
For Vela, analysed only in the earlier \citet{Keitel_2019} study
with different template placement,
the resulting \mbox{$\Ntemp \approx 3.6\times10^{5}$}
is lower than the $\sim 10^7$ used previously.

\vspace{-\baselineskip}

\subsection{Threshold estimation from mock search results}
\label{sec:depth-thresh}
The general idea is that we need to understand the noise-only distribution of our detection statistic
in order to determine how likely are outliers above a certain detection threshold
due to noise fluctuations alone,
i.e. false alarms.
This depends on the number of templates, because
more templates mean probing more independent noise realizations.
The approach in \citet{Tenorio:2021wad} is to empirically estimate 
the distribution of the \emph{loudest outlier} of a detection statistic 
from realizations of the distribution of the detection statistic itself
over a certain template bank.
This is implemented in the \texttt{distromax} package~\citep{distromax}.
Compared to previous approaches,
it still works if the distribution of the detection statistic
is not known in closed form
(as is the case for
the $\max \mathcal{F}$
and $\mathcal{B}_\mathrm{tS/G}$ statistics
for tCWs),
and also works robustly on banks with levels of correlation between templates
as used in practical searches
(non-vanishing mismatch between templates).

To create fake realizations of the distribution of the $\max \mathcal{F}$-statistic,
under the Gaussian noise assumption,
we use the \texttt{synthesizeTransientStats} program
from \texttt{LALSuite} \citep{Prix_G_M,LALSuite}.
Per realization, we draw $\Ntemp$ statistic values
corresponding to rectangular windowed tCWs over transient search windows of
$t_0$ in $T_\text{glitch}\pm1$\,day
and $\tau_\rwin \in [0,120]$\,days
with time resolutions of 1800\,s, 
and given the position of each pulsar.

We then determine a detection threshold using \texttt{distromax}.
Following the choice in \citet{our_paper},
the detection threshold corresponds to 
the mean of the estimated distribution of the loudest outlier
plus one standard deviation.
We have simulated four realizations of the noise distribution for each target,
with different random seeds, 
to control for the variance in estimated thresholds.
The $\max \mathcal{F}$ thresholds for Vela vary from 48.8 to 50.1,
so we use a rounded value of 50.
In the case of the Crab, thresholds vary from 51.5 to 52.8
and we use a rounded value of 52.

\vspace{-\baselineskip}

\subsection{Depth estimation}
\label{sec:depth-octapps}

The final step for the sensitivity depth estimation is based on
the procedure from \citet{Dreissigacker2018},
implemented in \texttt{OctApps} \citep{Wette_2018}.
This requires the following input parameters:
duty factor,
mismatch,
probability of false dismissal ($p_\mathrm{FD}$),
detection threshold,
sky location
and observation time $\tau_\rwin$.

Duty factors represent the fraction of time each detector
has been operative during the search.
In real observing runs to date, this factor can vary broadly
depending on the start time and duration of the analysis.
Here we choose a value of 0.66 per detector.
We further take the \citet{our_paper} setup as a reference to set values of
0.2 for the allowed mismatch and 0.05 for $p_\mathrm{FD}$.
The latter also corresponds to the typical UL confidence level.

Finally, we use the thresholds determined in \ref{sec:depth-thresh}, 
equivalent to setting a probability of false alarm,
and the Crab and Vela sky locations.
Then, using an observation time
\mbox{$\tau_\rwin = 10$\,days},
we obtain sensitivity depths of
\mbox{$\mathcal{D}_\rwin \approx 27.5$}
and $29.5/\sqrt{\mathrm{Hz}}$
for searches for Crab and Vela glitches respectively,
considering the two LIGO detectors.

For ET, the ASD we use~\citep{ET-0000A-18} in Section \ref{sec:5}
is that of a single detector,
which corresponds to the combination of one low- and one high-frequency interferometer
in the ET-D design~\citep{ET_sens_1}.
For the full observatory with its 3 detectors
and arm opening angles of 60 degrees,
an easy approach is to use a factor of
\mbox{$\sqrt{3}\sin(\pi/3) = 1.5$}
compared to a search with unit depth using a single L-shaped interferometer.
Instead, we have modified the \texttt{OctApps} code
to properly account for this geometry through the detector response,
with the parameters for a triangular ET at the Virgo site
as in \texttt{LALSuite} \citep{LALSuite},
which means that this factor does not have to be applied separately.
This yields sensitivity depths for Crab and Vela glitches
of \mbox{$\mathcal{D}_\rwin \approx 28.2$}
and $32.1/\sqrt{\mathrm{Hz}}$,
again at \mbox{$\tau_\rwin = 10$\,days}.

CE, on the other hand, has the same geometry as LIGO
even though its arms are longer.
Hence, the sensitivity depths for a single CE detector at Hanford
are the same as for LIGO Hanford alone,
i.e. for the same signal duration and Crab and Vela respectively, 
we get \mbox{$\mathcal{D}_\rwin \approx 18.4$}
and $21.6/\sqrt{\mathrm{Hz}}$.
The ASD we use in Section \ref{sec:5} corresponds to the ``second stage'' (CE2)
from \citet{CE_def}.
The reference concept in \citet{Evans:2021gyd} is one 40\,km
plus one 20\,km detector. 
With the \texttt{OctApps} sensitivity estimation code
we cannot directly model such a heterogeneous network, 
but its depth would be between the mean of these values
and that for the two LIGOs. 
Given the uncertainty in the actual network to be implemented, 
here we work with the single detector scenario
as a conservative assumption.

\vspace{-\baselineskip}

\subsection{Depth dependence on signal duration}
\label{sec:depth-taufit}
As discussed in Section~\ref{sec:5},
sensitivity depths for tCWs
(of either window shape)
are expected to scale with the square root of the signal duration parameter.
For the values just quoted, we have assumed \mbox{$\tau_\rwin = 10$\,days},
and most results in the main part of the paper
can be easily rescaled to different durations
or to exponential windows
as discussed in each case.
However, we have recomputed the mean of $\mathcal{D}_\rwin$
for searches for Crab and Vela glitches,
just like in step \ref{sec:depth-octapps},
with ET at 16 different values of $\tau_\rwin$ from 0.5 to 120 days,
and fit these values to produce a smooth curve in Fig. \ref{fig:eps_vs_tau},
obtaining in this case
$\mathcal{D}_\rwin \approx \sqrt{91 \,\tau_\rwin}$.

\bsp	
\label{lastpage}
\end{document}